\documentclass[journal]{IEEEtran}
\usepackage[table]{xcolor}
\usepackage{booktabs}
\usepackage{multirow}
\usepackage{color}
\usepackage{subcaption}
\usepackage{algorithm}
\usepackage{algorithmicx}
\usepackage{algpseudocode}
\usepackage{multirow}
\usepackage{cite}
\usepackage{mathtools}
\usepackage{bm}
\ifCLASSINFOpdf
   \usepackage[pdftex]{graphicx}
\else
\fi
\begin{document}
\title{Enhanced PeerHunter: Detecting Peer-to-peer Botnets through Network-Flow Level Community Behavior Analysis}
% \title{\huge Enhanced PeerHunter: Detecting Peer-to-peer Botnets through Network-Flow Level Community Behavior Analysis}

\author{Di~Zhuang,~\IEEEmembership{Student~Member,~IEEE,}
        and~J.~Morris~Chang,~\IEEEmembership{Senior~Member,~IEEE}% <-this % stops a space
\thanks{The authors are with the Department of Electrical Engineering, University of South Florida, Tampa, FL 33620 USA (e-mail: dizhuang@mail.usf.edu; chang5@usf.edu).}% <-this % stops a space
% \thanks{Manuscript received April 19, 2018; revised August 26, 2018.}
 }

\maketitle
\begin{abstract}
Peer-to-peer (P2P) botnets have become one of the major threats in network security for serving as the fundamental infrastructure for various cyber-crimes. More challenges are involved in the problem of detecting P2P botnets, despite a few work claimed to detect centralized botnets effectively. We propose Enhanced PeerHunter, a network-flow level community behavior analysis based system, to detect P2P botnets. Our system starts from a P2P network flow detection component. Then, it uses ``mutual contacts'' to cluster bots into communities. Finally, it uses network-flow level community behavior analysis to detect potential botnets. In the experimental evaluation, we propose two evasion attacks, where we assume the adversaries know our techniques in advance and attempt to evade our system by making the P2P bots mimic the behavior of legitimate P2P applications. Our results showed that Enhanced PeerHunter can obtain high detection rate with few false positives, and high robustness against the proposed attacks.
\end{abstract}
\IEEEpeerreviewmaketitle

\section{Introduction}
\label{sec:sec1}
\IEEEPARstart{A} botnet is a set of compromised machines controlled by botmasters through command and control (C\&C) channels. Botnets may have different communication architectures. Traditional botnets are known to use centralized architectures, which have potential single point of failure. Peer-to-peer (P2P) network is modeled as a distributed architecture, where even if a certain number of peers do not function properly, the whole network is not compromised. Most of the recent botnets (e.g., Storm, Waledac and ZeroAccess) attempted to use P2P architectures, and P2P botnets were proved to be highly resilient even after a certain number of bots being identified or taken down \cite{rossow2013sok}. P2P botnets provide a fundamental infrastructure for various cyber-crimes, such as distributed denial-of-service (DDoS), email spam, click fraud, etc. For instance, recent botnet attacks including those carried out by WhiskeyAlfa (responsible for Sony Pictures Entertainment attack) and WannaCry (responsible for ransoming healthcare facilities in Europe) showed the scale and scope of damage that P2P botnets can cause. As such, detecting P2P botnets effectively is rather important for securing cyberspace.

Designing an effective P2P botnets detection systems is very challenging.
First, botnets tend to act stealthily \cite{zhang2014building} and spend most of their time in the waiting stage before performing any malicious activities \cite{hang2013entelecheia}. Approaches using malicious activities would have small window of opportunities to detect such botnets. Second, botnets tend to encrypt the C\&C channels, causing deep-packet-inspection (DPI) based methods ineffective. Third, the role of a single bot can be changed dynamically depending on the current structure of a botnet \cite{yan2013peerclean} (e.g., P2P bot can shift its functionality to act as a botmaster when the prior botmaster has been taken down). Hence, it is difficult to characterize a botnet just by looking at a single bot.

In this work, we present Enhanced PeerHunter, an extension of PeerHunter \cite{zhuang2017peerhunter}, aiming to use network-flow level community behaviors to detect waiting stage P2P botnets, even in the scenario that P2P bots and legitimate P2P applications are running on the same set of hosts. We consider a botnet community as a group of compromised machines that communicate with each other or connect to the same set of botmasters through the same C\&C channel, are controlled by the same attacker, and aim to perform similar malicious activities. In the ``waiting stage'', no malicious activities could be observed. As discussed in \cite{yan2013peerclean}, the dynamic change of communication behaviors of P2P botnets makes it extremely hard to identify a single bot. Nonetheless, bots belonging to the same P2P botnet always operate together as a community and share the same set of community behaviors. Our system starts from a P2P network flow detection component, and builds a network-flow level mutual contacts graph (MCG) depending on the mutual contacts characteristics \cite{coskun2010friends} between each pair of the P2P network flows. Afterwards, it employs a community detection component to cluster the same type of bots into the same community, and separate bots and legitimate applications or different types of bots into different communities. Finally, our system uses the $destination \ diversity$ (the ``P2P behavior'') and the $mutual \ contacts$ (the ``botnet behavior'') as the natural behaviors to detect P2P botnet communities.

In the experiments, we mixed a background network dataset \cite{mawi} with 5 P2P botnets datasets and 4 legitimate P2P applications datasets \cite{rahbarinia2014peerrush}. To make our experimental evaluation as unbiased and challenging as possible, we propose a network traces sampling and mixing method to generate synthetic experimental datasets. To be specific, we evaluated our system with 100 synthetic experimental datasets that each contains 10,000 internal hosts. We implemented our P2P network flow detection component using MapReduce framework, which dramatically reduced the number of hosts subject to analysis by 99.03\% and retained most of the P2P hosts. Also, the MapReduce design and implementation of our system could be deployed on cloud-computing platforms (e.g., Amazon EC2), which ensures the scalability of our system (i.e., processing an average of 97 million network flows in about 20 minutes). To summarize, our work has the following contributions:

$\bullet$ We present a novel, effective and efficient network-flow level community behavior analysis based system, Enhanced PeerHunter, which is capable of detecting P2P botnets when (a) botnets are in their waiting stage; (b) the C\&C channel has been encrypted; (c) the botnet traffic are overlapped with legitimate P2P traffic on the same host; and (d) none statistical traffic patterns are known in advance (unsupervised).

$\bullet$ We experimented our system using a wide range of parameter settings. With the best parameter settings, our system achieved 100\% detection rate with zero false positive.

$\bullet$ We propose two evasion attacks (i.e., passive and active mimicking legitimate P2P application attacks), where we assume the adversaries know our techniques in advance and attempt to evade our system via instructing P2P bots to mimic the behavior of legitimate P2P applications. The experiment results showed that our system is robust to both attacks.

$\bullet$ We compared Enhanced PeerHunter with PeerHunter \cite{zhuang2017peerhunter} (i.e., our previous work) and Zhang \textit{et al.} \cite{zhang2014building}. Extensive experiments were conducted to show that (a) our system outperforms Zhang \textit{et al.} \cite{zhang2014building} in terms of the detection rate of different botnets, the overall precision, recall and false positives, and (b) our system is more robust to MMKL attacks compared with PeerHunter \cite{zhuang2017peerhunter} and Zhang \textit{et al.} \cite{zhang2014building}.

The rest of this paper is organized as follows:
Section~\ref{sec:sec2} presents the related work.
Section~\ref{sec:sec3} explains the motivation and details of the features applied in our system.
Section~\ref{sec:sec4} describes the system design and implementation details.
Section~\ref{sec:sec5} presents the experimental evaluation.
Section~\ref{sec:sec8} discusses the evasions and possible solutions, deployment and the potential extensions of our system.
Section~\ref{sec:sec7} concludes.

\section{Related Work}
\label{sec:sec2}
To date, a few methods attempting to detect P2P botnets were proposed \cite{gu2007bothunter, coskun2010friends, nagaraja2010botgrep, hang2013entelecheia, yan2013peerclean, rahbarinia2014peerrush, zhang2014building, zhuang2017peerhunter, wang2017botnet, venkatesan2017detecting, karuppayah2017sensorbuster, haddadi2017botnet}. From the data perspective, recent approaches can be divided into two categories \cite{haddadi2017botnet}: payload-based and flow-based. Payload-based systems \cite{gu2007bothunter, lu2011clustering, wang2018detecting} use payload content and header information of network packets to detect botnets. For instance, BotHunter \cite{gu2007bothunter} is a well-known packet inspecting bot detection system that relies on a modified Snort \cite{roesch1999snort} (i.e., a rule-based intrusion detection system that requires the  access to the full payload) to detect potential malicious activities and further identify infected hosts. Lu \textit{et al.} \cite{lu2011clustering} proposed to use decision tree models trained on the n-gram features extracted from the network traffic payload to detect botnets. Wang \textit{et al.} \cite{wang2018detecting} proposed to use lexical features of HTTP header (TCP payload) to discover malicious behaviors of Android botnets.

Flow-based systems \cite{coskun2010friends, nagaraja2010botgrep, ma2010novel, hang2013entelecheia, yan2013peerclean, rahbarinia2014peerrush, zhang2014building, zhuang2017peerhunter, wang2017botnet, venkatesan2017detecting, karuppayah2017sensorbuster, haddadi2017botnet, khanchi2018botnet} use header information of network packets (i.e., network flow characteristics) to capture botnets behaviors. Compared with payload-based systems, flow-based systems use less information from the network packets. Since recent botnets tend to use encryption to hide their payload information from the detection systems, most of the packet-based systems that applying deep packet inspection (DPI) on the payload information (e.g., BotHunter \cite{gu2007bothunter}) will be foiled. Zhang \textit{et al.} \cite{zhang2013detecting} proposed to add a high-entropy flow detector into BotHunter to detect bots, when part of the packets payloads of botnets' network flows are encrypted. Their assumption is that the presence of high-entropy flows (detected from the encrypted packets payloads) together with existing botnets events (detected from the non-encrypted packets payloads by BotHunter) could identify botnets using encrypted network traffic. However, if all the packets payloads are encrypted \cite{haddadi2017botnet}, it will be hard for their approach to perform. The flow-based detection systems have advantage over the packet-based systems that applying deep packet inspection (DPI) on the payload information (e.g., BotHunter \cite{gu2007bothunter}) given that they can be applied to encrypted traffic. Some flow-based systems applied one or several different supervised machine learning algorithms on a set of well extracted network flow features to model the botnets behaviors. For instance, Jianguo \textit{et al.} \cite{jianguo2016botnet} applied three supervised machine learning algorithms (i.e., SVM, Logistic Regression and Neural Network) on network flow features extracted from Netmate and Tranalyzer to detect botnets. They obtained very high performance metrics, while employing a fully labelled dataset.
Khanchi \textit{et al.} \cite{khanchi2018botnet} proposed an approach using genetic programming and ML on data streams to detect botnets flows. However, since most of the supervised ML-based approaches usually generate models that are focusing on specific types of botnets (existing in the training data), those approaches will not be effective to detect botnets not appeared in the training data (unknown botnets).

Some flow-based systems utilized a combination of different heuristics to model P2P botnets behaviors. For instance,
Botgrep \cite{nagaraja2010botgrep} proposed to detect P2P botnets through localizing structured communication graphs, where they found that the communication graph of P2P applications have fast convergence time of random walks to a stationary distribution. However, their method can only identify structured communication subgraphs, rather than ensure those subgraphs containing P2P botnets.
Entelecheia \cite{hang2013entelecheia} proposed to use a synergistic graph-mining approach on a super-flow graph built from network flow features (i.e., volume per hour, duration per flow) to identify a group of P2P bots, where they claimed that P2P botnet network flow tend to have low volume and long duration.
Group or community behavior based methods \cite{yan2013peerclean, coskun2010friends, zhuang2017peerhunter, wang2017botnet} considered the behavior patterns of a group of bots within the same P2P botnet community. Coskun \textit{et al.} \cite{coskun2010friends} developed a P2P botnets detection approach that started from building a mutual contacts graph of the whole network, then attempted to use ``seeds'' (known bots) to identify the rest of botnets. However, it is impractical to have a ``seed'' in advance. Similar to the idea of using mutual contacts graph, Ma \textit{et al.} \cite{ma2014dnsradar} proposed to use the coexistence of domain cache-footprints distributed in networks that participate in the outsourcing service (i.e., coexistence graph) to detect malicious domains.
Yan \textit{et al.} \cite{yan2013peerclean} proposed a group-level behavior analysis based P2P botnets detection method, where they started from clustering P2P hosts into groups, and then used supervised machine learning methods (e.g., SVM) to identify bots through a set of group-level behavior features. Since their approach relied on supervised classification methods (e.g., SVM) which required to train the model of each botnet on fully labelled dataset in advance, it would be hard for their method to detect unknown botnets.
Chen \textit{et al.} \cite{chen2017exploring} applied three unsupervised machine learning algorithms (i.e., self-organising map, local outlier factor and k-NN outlier) to build a normal behavior profile to detect botnet. They obtained a very high detection rate (91.3\%), but with inherited high false positive rates due to the nature of the unsupervised ML algorithms employed.
PeerHunter \cite{zhuang2017peerhunter}, our previous work, proposed to use the host level community behavior analysis to detect P2P botnets, which did not consider the scenario that P2P bots and legitimate P2P applications could run on the same set of hosts. Zhang \textit{et al.} \cite{zhang2014building} proposed a scalable botnet detection system capable of detecting stealthy P2P botnets (i.e., in the waiting stage), where no knowledge of existing malicious behavior was required in advance. They also claimed to work in the scenario that the botnet traffic are overlapped with the legitimate P2P traffic on the same host. However, their experimental dataset was slightly biased and less challenging. For example, in their dataset, the number of bots was twice as many as the number of legitimate P2P hosts, which was much easier for bots to form clusters than legitimate P2P hosts.

In this work, we present Enhanced PeerHunter, a network-level flow-based system that relies on community behavior analysis to detect P2P botnets. We compared Enhanced PeerHunter with PeerHunter \cite{zhuang2017peerhunter} and Zhang \textit{et al.} \cite{zhang2014building} on a more challenging and comprehensive experimental datasets, and showed that our system outperforms both systems in terms of detection rate, false positives and the performance under the proposed mimicking legitimate P2P application attacks.

\section{Background and Motivation}
\label{sec:sec3}
In this section, we investigate the characteristics being used to detect P2P network traffic, and introduce the concept of ``mutual contacts'', which motivated us to formulate the P2P botnet detection problem as a network community detection problem. Also, we explore the P2P botnet community behaviors being used to identify botnets communities.
To demonstrate the features discussed in this section, we conducted some preliminary experiments using the dataset shown in Table~\ref{table:D1} and Table~\ref{table:D2}. Table~\ref{table:notations} shows the notations and descriptions, and Table~\ref{table:prelim} shows the measurements of features.
\begin{table}[h]
\footnotesize
\captionsetup{font=footnotesize}
\caption{Notations and Descriptions}
\label{table:notations}
\centering
\begin{tabular}{l|l}
\hline
\bfseries Notations & \bfseries Descriptions\\
\hline
MNF & the management network flow\\
\hline
AVGDD & the average \# of distinct /16 MNF dstIP prefixes\\
\hline
AVGDDR & the average destination diversity ratio\\
\hline
AVGMC & the average \# of mutual contacts between a pair of hosts\\
\hline
AVGMCR & the average mutual contacts ratio\\
\hline
$\Theta_{dd}$ & the threshold of destination diversity\\
\hline
$\Theta_{mcr}$ & the threshold of mutual contacts ratio\\
\hline
$\Theta_{avgddr}$ & the threshold of AVGDDR\\
\hline
$\Theta_{avgmcr}$ & the threshold of AVGMCR\\
\hline
BSI & Bot Separation Index\\
\hline
BAI & Bot Aggregation Index\\
\hline
BLSI & Bot-Legitimate Separation Index\\
\hline
\end{tabular}
\end{table}

\begin{table}[h]
\footnotesize
\captionsetup{font=footnotesize}
\caption{Measurements of Features}
\label{table:prelim}
\centering
\begin{tabular}{l|r|r|r|r}
\hline
\bfseries Trace & \bfseries AVGDD & \bfseries AVGDDR  & \bfseries AVGMC & \bfseries AVGMCR \\
\hline
  eMule & 8,349 & 17.6\% & 3,380 & 3.7\%\\
  \hline
  FrostWire & 11,420 & 15.2\% & 7,134 & 4.5\%\\
  \hline
  uTorrent & 17,160 & 8.7\% & 13,888 & 3.5\%\\
  \hline
  Vuze & 12,983 & 10.1\% & 18,850 & 7.9\%\\
  \hline
  Storm & 7,760 & 25.1\% & 14,684 & 30.2\%\\
  \hline
  Waledac & 6,038 & 46.0\% & 7,099 & 37.0\%\\
  \hline
  Sality & 9,803 & 9.5\% & 72,495 & 53.2\%\\
  \hline
  Kelihos & 305 & 97.4\% & 310 & 98.2\%\\
  \hline
  ZeroAccess & 246 & 96.9\% & 254 & 100.0\%\\
\hline
\end{tabular}
\end{table}
\subsection{P2P Network Characteristics}
\label{sec:sec3_1}
Due to the nature of P2P networks, P2P hosts usually communicate with their peers through IP addresses directly, without any queries from DNS services \cite{wu2009identifying}, namely, non-DNS connections (NoDNS). Also, peer churn is another typical behavior in P2P networks \cite{stutzbach2006understanding}, which results in a significant number of failed connections in P2P network flow. Furthermore, due to the decentralized nature of P2P network, a P2P host usually communicates with peers distributed in a large range of physical networks, which results in destination diversity (DD) \cite{rahbarinia2014peerrush} of P2P management network flow (MNF).
To be clearer, P2P host generate two types of network flow: (1) management network flow, which maintains the function and structure of the P2P network, and (2) other network flow, such as data-transfer flow, which does not necessarily have the P2P network characteristics. The P2P network flow mentioned in this section and the rest all refers to P2P MNF.

Zhang \textit{et al.} \cite{zhang2014building} proposed to remove a decent number of non-P2P network flow using NoDNS, and then performed a fine-grained P2P hosts detection using DD. Based on their experiment results, DD plays a much more important role in detecting P2P hosts than NoDNS. Therefore, in this work, we decided to only use DD to simplify and speed up the P2P network flow detection procedure.
In addition, we used the number of distinct /16 IP prefixes of each host's network flow, rather than BGP prefix used in \cite{zhang2014building} to approximate DD, since /16 IP prefix is a good approximation of network boundaries. For instance, it is very likely that two IP addresses with different /16 IP prefixes belong to two distinct physical networks. This is also supported by Table~\ref{table:prelim}, which shows the network flow in a P2P network spreading across many distinct physical networks according to the number of /16 IP prefixes.

\subsection{Mutual Contacts}
\label{sec:sec3_2}
\begin{figure}[tb]
\captionsetup{font=footnotesize}
        \centering
        \begin{subfigure}[b]{0.34\textwidth}
                \includegraphics[width=\textwidth]{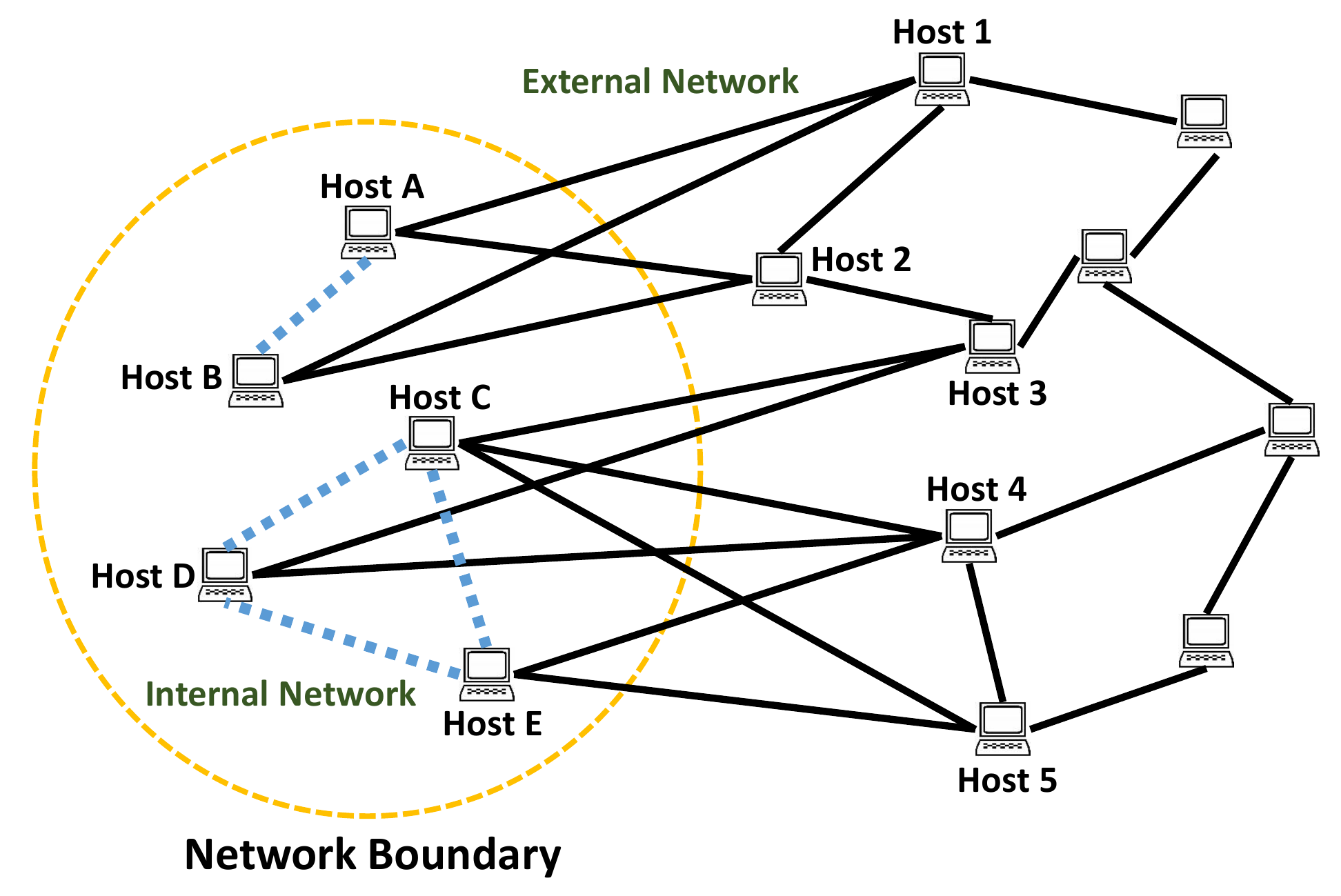}
                \caption{}
                \label{fig:network1}
        \end{subfigure}%
        ~ %add desired spacing between images, e. g. ~, \quad, \qquad, \hfill etc.
        \begin{subfigure}[b]{0.14\textwidth}
                \includegraphics[width=\textwidth]{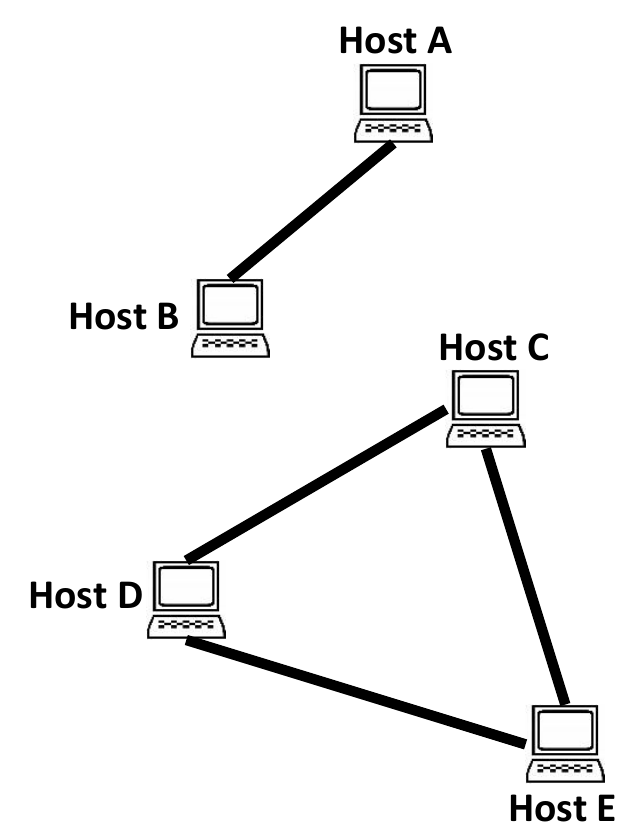}
                \caption{}
                \label{fig:network2}
        \end{subfigure}
        \caption{Illustration of network (a) and its mutual contacts graph (b).}
        \label{fig:MCG}
\end{figure}

The mutual contacts (MC) between a pair of hosts is a set of shared contacts between them \cite{coskun2010friends}. Consider the network illustrated in Fig.~\ref{fig:network1} which contains an internal network (A, B, C, D and E) and an external network (1, 2, 3, 4 and 5). A link between a pair of hosts means communication between them.
In Fig.~\ref{fig:network1}, 1, 2 are the mutual contacts shared by A, B.

Mutual contacts are the natural characteristic of P2P botnet. Compared with legitimate hosts, a pair of bots within the same P2P botnet has higher probability to share mutual contacts \cite{coskun2010friends}. Because bots within the same P2P botnet tend to receive the same C\&C messages from the same set of botmasters \cite{holz2008measurements}. Moreover, in order to prevent bots (peers) from churning, the botmaster must check each bot periodically, which results in a convergence of contacts among peers within the same botnet \cite{zhang2014building}. However, since bots from different botnets are controlled by different botmasters, they will not share many mutual contacts. A pair of Legitimate hosts may have a small set of mutual contacts, since nearly all hosts communicate with some popular servers, such as google.com, facebook.com \cite{coskun2010friends}. Furthermore, the host pairs running the same P2P applications may also result in a decent ratio of mutual contacts, if they communicate with the same set of peers by coincidence. However, in practice, legitimate P2P hosts with different purposes will not search for the same set of peers. As such, we can use mutual contacts to cluster the bots within the same botnet, and separate P2P botnets from legitimate P2P applications.

The basic idea of using mutual contacts is to build a mutual contacts graph (MCG) as shown in Fig.~\ref{fig:MCG}, a host level MCG, where A, B are linked together in Fig.~\ref{fig:network2}, since they have mutual contacts 1, 2 in Fig.~\ref{fig:network1}. Similarly, C, D, E are linked to each other in Fig.~\ref{fig:network2}, since every pair of them share at least one mutual contacts in Fig.~\ref{fig:network1}. More details about network-flow level MCG is discussed in Section~\ref{sec:sec4_3}.

\subsection{Community Behavior Analysis}
\label{sec:sec3_3}
Due to the dynamic changes of a single bot's communication behavior \cite{yan2013peerclean}, it would be extremely hard to identify a single bot. However, bots within the same P2P botnet always work together as a community, thus, should have distinguishable community behaviors. We consider three types of community behaviors: (a) flow statistical feature, (b) numerical community feature and (c) structural community feature.

\subsubsection{Flow Statistical Feature}
\label{sec:sec3_3_1}
Botnet detection methods using flow statistical features, have been widely discussed \cite{yan2013peerclean, hang2013entelecheia, zhang2014building, zhuang2017peerhunter}. For the MNFs of a specific P2P application, most of its statistical patterns depend on its P2P network protocol. However, the statistical patterns of other network flows, such as data-transfer flow, are usually situation-dependent, which vary a lot even in the same P2P network. In this work, we use the ingoing and outgoing bytes-per-packets (BPP) of MNFs in one P2P network as its community flow statistical feature.

\subsubsection{Numerical Community Feature}
\label{sec:sec3_3_2}
We consider two numerical community features: average destination diversity ratio (AVGDDR) and average mutual contacts ratio (AVGMCR).

{\bf Average Destination Diversity Ratio: }
This captures the ``P2P behavior'' of P2P botnets. The destination diversity (DD) of a P2P host is the number of distinct /16 IP prefixes of its network flows' destination IPs. The destination diversity ratio (DDR) of each host is its DD divided by the total number of distinct destination IPs of its network flows.
Due to the decentralized nature of P2P networks, P2P network flow tend to have higher DDR than non-P2P network flow. Furthermore, network flow from P2P botnets usually have higher AVGDDR than network flow from legitimate networks. Network flow from bots within the same botnet tend to have similar DDR, since those bots are usually controlled by machines, rather than humans. However, the destinations of legitimate P2P network flow are usually user-dependent, which result in their DDR varying greatly from user to user. Besides, our approach aims to cluster bots within the same botnets together, rather than attempting to cluster the legitimate hosts. Therefore, legitimate communities might contain both P2P hosts and non-P2P hosts, leading to lower AVGDDR. As shown in Table~\ref{table:prelim}, both legitimate hosts and bots spread across a wide range of distinct networks. However, most of the botnets have higher AVGDDR than legitimate applications, except Sality. % We could combine the next feature to identify Sality.

{\bf Average Mutual Contacts Ratio: }
This captures the ``botnet behavior'' of P2P botnets. The mutual contacts ratio (MCR) between a pair of hosts is the number of mutual contacts between them, divided by the number of total distinct contacts of them. This is based on three observations: (a) P2P botnets are usually formed by at least two bots, otherwise they cannot act as a group, (b) the MCR of a pair of bots within the same botnet is much higher than the MCR of a pair of legitimate applications or a pair of bots from different botnets, and (c) each pair of bots within the same botnet has similar MCR. Thus, we define AVGMCR as the average MCR among all pairs of hosts within one network community. As shown in Table~\ref{table:prelim} both botnets and certain legitimate network communities have a considerable number of mutual contacts. That is because those legitimate communities have much more ``base'' contacts than botnets. However, botnets have much higher AVGMCR.

\subsubsection{Structural Community Feature}
\label{sec:sec3_3_3}
This captures the structural characteristics of a botnet. As discussed above, every pair of bots within the same botnet tends to have a considerable number or ratio of mutual contacts. If we consider each host as a vertex and link an edge between a pair of hosts when they have mutual contacts, the bots within the same botnet tend to form cliques. On the contrary, the contacts of different legitimate hosts usually diverge into different physical networks. Thus, the probability that legitimate communities form certain cliques is relatively low. Then, we can consider P2P botnets detection as a clique detection problem, which detects cliques from a given network with certain requirements. However, since clique detection problem is NP-complete, we cannot directly apply such method to detect botnets, without any preprocessing. We propose to combine all three botnet community behaviors, and use the previous two community behaviors as the ``preprocessing'' of the clique detection problem.

\begin{figure*}[tb]
\centering
\includegraphics[width=475pt]{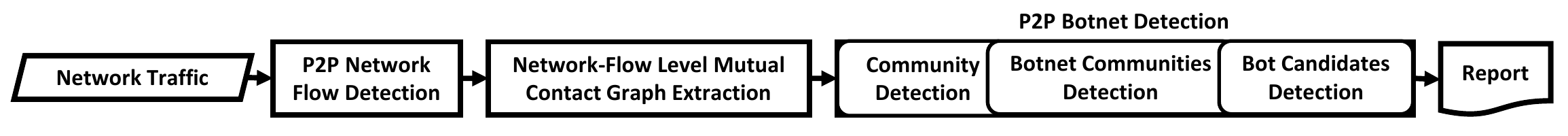}
\caption{System Overview.}
\label{fig:sysview}
\end{figure*}

\section{System Design}
\label{sec:sec4}
Enhanced PeerHunter has three components, as shown in Fig.~\ref{fig:sysview}, that work synergistically to (a) detect P2P network flow, (b) construct the network-flow level mutual contacts graph, and (c) detect P2P botnets.

\subsection{P2P Network Flow Detection}
\label{sec:sec4_2}
This component aims to detect network flow that engage in P2P communications using the features described in Section~\ref{sec:sec3_1}. The input is a set of 5-tuple network flow [$ip_{src}$, $ip_{dst}$, $proto$, $bpp_{out}$, $bpp_{in}$], where $ip_{src}$ is the source IP, $ip_{dst}$ is the destination IP, $proto$ is either tcp or udp, and $bpp_{out}$ and $bpp_{in}$ are outgoing and ingoing bytes-per-packets (BPP) statistics. First, we group all network flows $F=\{f_{1}$, $f_{2}$, $\dots$, $f_{k}\}$ into flow clusters $FC=\{FC_{1}$, $FC_{2}$, $\dots$, $FC_{m}\}$ using the 4-tuple [$ip_{src}$, $proto$, $bpp_{out}$, $bpp_{in}$]. Then, we calculate the number of distinct /16 prefixes of $ip_{dst}$ (destination diversity) associated with each flow cluster, $dd_{i}=DD(FC_{i})$. If $dd_{i}$ is greater than a pre-defined threshold $\Theta_{dd}$, we consider $FC_{i}$ as a P2P MNF cluster, and its source hosts as P2P hosts. We retain all the network flows within the P2P MNF clusters for the next component, and eliminate all the other network flows.
\begin{algorithm}
\caption{P2P Network Flow Detection}\label{alg:alg1}
\begin{algorithmic}[1]
\Function{Map}{$[ip_{src}, ip_{dst}, proto, bpp_{out}, bpp_{in}]$}
\State $Key \gets [ip_{src}, proto, bpp_{out}, bpp_{in}]$
\State $Value \gets ip_{dst}$
\State \textbf{output} $(Key, Value)$
\EndFunction
\Function{Reduce}{$Key$, $Value[$ $]$}
\State $k \gets Key$
\State $dd_{k}=\O$
\For {$v \in Value[$ $]$}
    \State $dd_{k} \gets dd_{k} \cup \{v\}$
\EndFor
\If {$|dd_{k}| \geq \Theta_{dd}$}
\For {$v \in Value[$ $]$}
    \State \textbf{output} $(k, v)$
\EndFor
\EndIf
\EndFunction
\end{algorithmic}
\end{algorithm}
As shown in Algorithm~\ref{alg:alg1}, we designed this component using a MapReduce framework \cite{dean2008mapreduce}. For a mapper, the input is a set of 5-tuple network flow, and the output is a set of key-value pairs, where the key is the 4-tuple [$ip_{src}$, $proto$, $bpp_{out}$, $bpp_{in}$], and the value is its corresponding $ip_{dst}$. For a reducer, the input is the set of key-values pairs that outputs by the mapper. Then, the reducer aggregates all values with the same key to calculate the DD of each flow cluster, and finally output the detected P2P MNF based on $\Theta_{dd}$.

\begin{figure*}[tb]
\centering
\includegraphics[width=500pt]{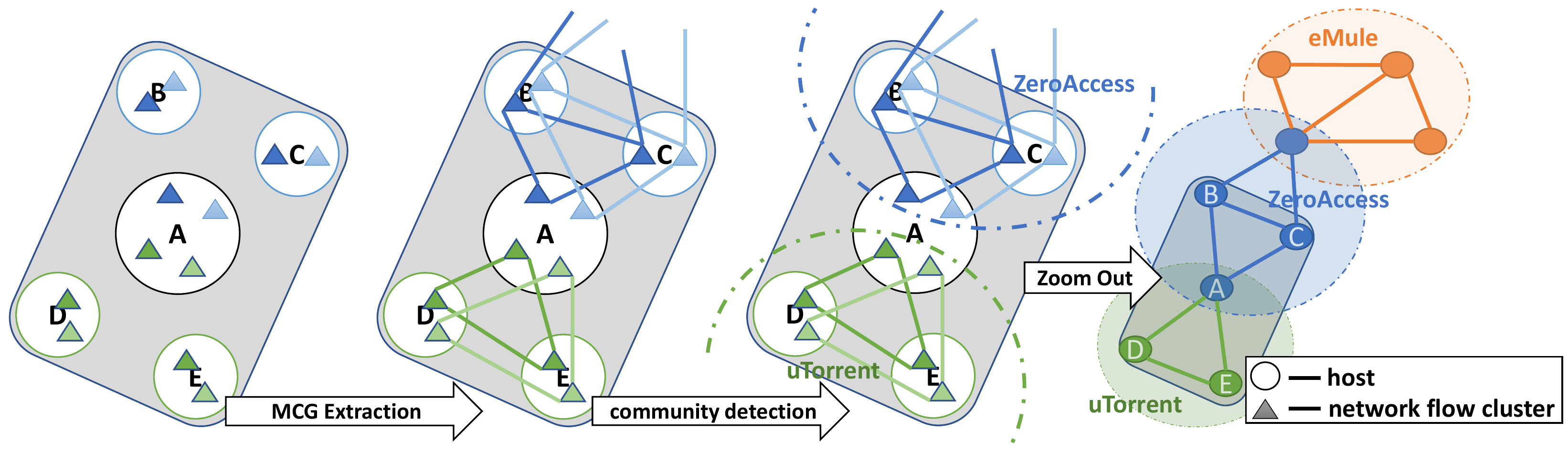}
\caption{An example of network-flow level mutual contacts graph extraction and community detection. Each triangle represents a network flow cluster, and the same color triangles represent the same type of network flow clusters. The areas separated by the dash-dot line with different color represents different communities.}
\label{fig:An_example}
\end{figure*}

\subsection{Network-Flow Level Mutual Contacts Graph Extraction}
\label{sec:sec4_3}
This component aims to extract mutual contacts graph (MCG) using the network-flow level mutual contacts. We call a pair of P2P network flow clusters are the same type, if they have the same 3-tuple [$proto$, $bpp_{out}$, $bpp_{in}$]. As illustrated in Fig.~\ref{fig:An_example}, each host might contain one type or several different types of P2P network flow clusters generated by either P2P botnets or legitimate P2P applications running on it. If a pair of the same type of P2P network flow clusters generated by different hosts, have at least one (network-flow level) mutual contacts, we create an edge between them in the corresponding network-flow level MCG.

To be specific, the input is a set of P2P network flow clusters $FC$=$\{FC_{1}$, $FC_{2}$, $\dots$, $FC_{m}\}$, and their corresponding P2P network flows, $F$=$\{f_{1}^1$, $f_{2}^1$, $\ldots$, $f_{n_{1}}^1$, $f_{1}^2$, $f_{2}^2$, $\ldots$, $f_{n_{2}}^2$, $\ldots$, $f_{1}^{|FC|}$, $f_{2}^{|FC|}$, $\ldots$, $f_{n_{|FC|}}^{|FC|} \}$, where $f_{i}^j$ denotes the flow $i$ of $FC_{j}$. The output is a MCG, $G_{mc} = (V, E)$, where each vertex $v_{i} \in V$ represents network flow cluster $FC_{i}$ and has a DDR score $ddr_{i}$, and each edge $e_{ij} \in E$ represents the existence of mutual contacts between $FC_{i}$ and $FC_{j}$ and has a nonnegative MCR weight $mcr_{ij}$. Algorithm~\ref{alg:alg2} shows the detailed steps.

First, for each P2P network flow cluster $FC_{i}$, we generate a contact set $C_{i}$, that contains all the destination IPs of its network flows. Each P2P network flow cluster $FC_{i}$ also contains a flow statistical pattern set $S_{i}$, which contains all the 3-tuple [$proto$, $bpp_{out}$, $bpp_{in}$] of its network flows. Let $DD(C_{i})$ be the set of distinct /16 prefixes of all the IPs in $C_{i}$. Then, $ddr_{i}$ and $mcr_{ij}$ can be calculated as follows.
\begin{equation}
    \label{eq:ddr}
    ddr_{i} = \frac{\| DD(C_{i}) \|}{\| C_{i} \|}   \ \ \ \ \ mcr_{ij} = \frac{\| C_{i} \cap C_{j} \|}{\| C_{i} \cup C_{j} \|}
\end{equation}

Furthermore, as discussed in Section~\ref{sec:sec3_3_1}, the network flows from different hosts (or network flow clusters) within the same network communities (generated by the same type of P2P botnet or legitimate P2P application) should have similar statistical patterns. Thus, for each pair of input P2P network flow clusters, say $FC_{i}$ and $FC_{j}$, we calculate the intersection between $S_{i}$ and $S_{j}$. If $S_{i} \cap S_{j} = \O$, then there should be no edge between $FC_{i}$ and $FC_{j}$ in MCG. Otherwise, they share at least one network flow statistical pattern, and we calculate $mcr_{ij}$ as shown in equation~(\ref{eq:ddr}). Let $\Theta_{mcr}$ be a pre-defined threshold. Then, if $mcr_{ij} > \Theta_{mcr}$, there is an edge between $FC_{i}$ and $FC_{j}$, with weight $mcr_{ij}$. Otherwise, there is no edge between $FC_{i}$ and $FC_{j}$ (i.e., $mcr_{ij}=0$).
\begin{algorithm}
\caption{Network-Flow Level MCG Extraction}\label{alg:alg2}
\begin{algorithmic}[1]
\renewcommand{\algorithmicrequire}{\textbf{input:}}
\renewcommand{\algorithmicensure}{\textbf{output:}}
\Require $FC$, $F$, $\Theta_{mcr}$
\Ensure $G_{mc} = (V, E)$

\State $E=\O$, $V=\O$
\For {$FC_{i} \in FC$}
    \State   $C_{i}=\O$
    \State   $S_{i}=\O$
\EndFor

\For {$f_{i}^j \in F$}
    \State $C_{j} \gets C_{j} \cup \{ip_{dst}\}$
    \State $S_{j} \gets S_{j} \cup \{[proto, bpp_{out}, bpp_{in}]\}$
\EndFor

\For {$FC_{i} \in FC$}
    \State $ddr_{i} \gets \frac{\| DD(C_{i}) \|}{\| C_{i} \|}$
    \State $vertex$ $v_{i} \gets <ddr_{i}>$
    \State $V \gets V \cup \{v_{i}\}$
\EndFor

\For {$\forall$ $FC_{i}, FC_{j} \in FC$ and $i<j$}
\If {$S_{i} \cap S_{j} \neq \O$}
    \State $mcr_{ij} \gets \frac{\| C_{i} \cap C_{j} \|}{\| C_{i} \cup C_{j} \|}$.
    \If {$mcr_{ij} > \Theta_{mcr}$}
        \State $edge$ $e_{ij} \gets <mcr_{ij}>$
        \State $E \gets E \cup \{e_{ij}\}$
    \EndIf
\EndIf
\EndFor

\State \Return $G_{mc} = (V, E)$
\end{algorithmic}
\end{algorithm}

\subsection{P2P Botnet Detection}
\label{sec:sec4_4}
This component aims to detect P2P bots from given MCG. First, we cluster the bots and the other hosts into their own communities using a community detection method. Afterwards, we detect botnet communities using numerical community behavior analysis. Finally, we use structural community behavior analysis to further identify or verify each bot candidate. Algorithm~\ref{alg:alg3} shows the detailed steps.

\subsubsection{Community Detection}
Given MCG $G_{mc} = (V, E)$, $\forall$ $e_{ij} \in E$, we have $mcr_{ij} \in [0.0, 1.0]$, where $mcr_{ij}=1.0$ means all contacts of $FC_{i}$ and $FC_{j}$ are mutual contacts and $mcr_{ij}=0.0$ means there is no mutual contact between $FC_{i}$ and $FC_{j}$.
Furthermore, the same type of P2P network flow clusters that generated by different bots within the same botnet tend to have a higher ratio of mutual contacts.
As such, the P2P bots clustering problem can be considered as a network community detection problem.
As shown in Fig.~\ref{fig:An_example}, each host might be running P2P bots or legitimate P2P applications or both, and each P2P bot or each legitimate P2P application generates different types of network flow clusters. Our community detection aims to cluster the same type of P2P network flow clusters generated by different bots into the same network flow cluster community. As such, each network flow cluster should only belong to a single network flow cluster community, but each host might belong to different host communities. Also, each botnet might contain several different network flow cluster communities. Once one network flow cluster community has been detected as belonging to a botnet, we consider the corresponding hosts as bots.

We used Louvain method, a modularity-based community detection algorithm \cite{blondel2008fast}, due to (a) its definition of a good community detection result (high density of weighted edges within communities and low density of weighted edges between communities) is perfect-suited for our P2P botnet community detection problem; (b) it outperforms many other modularity methods in terms of computation time \cite{blondel2008fast}; and (c) it can handle large network data sets (e.g., the analysis of a typical network of 2 million nodes takes 2 minutes \cite{blondel2008fast}).

Given $G_{mc}$ $=$ $(V,$ $E)$ as input, Louvain method outputs a set of network flow cluster communities $Com=\{com_{1},$ $com_{2},$ $\ldots,$ $com_{|Com|}\}$, where $com_{i}=(V_{com_{i}},$ $E_{com_{i}})$. $V_{com_{i}}$ is a set of network flow clusters in $com_{i}$. $E_{com_{i}}$ is a set of edges, where $\forall$ $e_{jk} \in E_{com_{i}}$, we have $e_{jk} \in E$ and $v_{j}, v_{k} \in V_{com_{i}}$.

\subsubsection{Botnet Communities Detection}
Given a set of communities $Com$, for each community $com_{i} \in Com$, we calculate its $avgddr_{i}$ and $avgmcr_{i}$ as follows.
\begin{equation}
    \label{eq:avgddr}
    avgddr_{i} = \frac{\sum_{v_{j} \in V_{com_{i}}} ddr_{j}}{\|V_{com_{i}}\|}
\end{equation}
\begin{equation}
    \label{eq:avgmcr}
    avgmcr_{i} = \frac{2 \times \sum_{\forall e_{jk} \in E_{com_{i}}} mcr_{jk}}{\|V_{com_{i}}\| \times (\|V_{com_{i}}\| - 1)}
\end{equation}

We define two thresholds $\Theta_{avgddr}$ and $\Theta_{avgmcr}$. 
$\forall$ $com_{i} \in Com$, if $avgddr_{i} \geq \Theta_{avgddr}$ and $avgmcr_{i} \geq \Theta_{avgmcr}$, we consider $com_{i}$ as a botnet network flow cluster community.

\subsubsection{Bot Candidates Detection}
Recall from Section~\ref{sec:sec3_3_3}, the MCG of botnet communities usually have a structure of one or several cliques. Therefore, we used a maximum clique detection method $CliqueDetection$ to verify each bot network flow cluster from botnet network flow cluster communities, and further identify bot candidates. Each time it tries to detect one or several maximum cliques on the given botnet (network flow cluster) communities. If the maximum clique (at least containing 3 vertices) has been found, we consider the network flow clusters in that clique as bot network flow cluster, and run the maximum clique detection algorithm on the remaining parts, until no more qualified maximum cliques to be found. Afterwards, we report the corresponding source hosts of the identified bot network flow clusters as the bot candidates.
\begin{algorithm}
\caption{P2P Botnet Detection}\label{alg:alg3}
\begin{algorithmic}[1]
\renewcommand{\algorithmicrequire}{\textbf{input:}}
\renewcommand{\algorithmicensure}{\textbf{output:}}
\Require $G_{mc}$, $\Theta_{avgddr}$, $\Theta_{avgmcr}$
\Ensure $S_{bot}$

\State $S_{botFCCom}=\O$, $S_{botFC}=\O$, $S_{bot}=\O$
\State $Com \gets {\bf Louvain}(G_{mc})$

\For {$com_{i} \in Com$}
    \State $avgddr_{i} \gets \frac{\sum_{v_{j} \in V_{com_{i}}} ddr_{j}}{\|V_{com_{i}}\|}$
    \State $avgmcr_{i} \gets \frac{2 \times \sum_{\forall e_{jk} \in E_{com_{i}}} mcr_{jk}}{\|V_{com_{i}}\| \times (\|V_{com_{i}}\| - 1)}$
    \If {$avgddr_{i} \geq \Theta_{avgddr}$ and $avgmcr_{i} \geq \Theta_{avgmcr}$}
        \State $S_{botFCCom} \gets S_{botFCCom} \cup \{com_{i}\}$
    \EndIf
\EndFor

\For {$com_{i} \in S_{botFCCom}$}
    \State $S_{botFC} \gets {\bf CliqueDetection}(com_{i})$
    \For {$FC_{i} \in S_{botFC}$}
        \For {$f^{i}_{j} \in FC_{i}$}
            \State $S_{bot} \gets S_{bot} \cup \{ip_{src}\}$
        \EndFor
    \EndFor
\EndFor

\State \Return $S_{bot}$

\end{algorithmic}
\end{algorithm}

\section{Experimental Evaluation}
\label{sec:sec5}
\subsection{Experiment Setup}
\label{sec:sec5_1}
\subsubsection{Experiment Environment}
\label{sec:sec5_1_1}
All the experiments were conducted on a PC with an 8 core Intel i7-4770 Processor, 32GB RAM, running 64-bit Ubuntu 16.04 LTS operating system. Our system was implemented using Java with JDK 8.

\subsubsection{Data Collection and Analysis Tool}
\label{sec:sec5_1_2}
We used three main datasets: (a) 24 hours network traces of 4 popular legitimate P2P applications, (b) 24 hours network traces of 5 P2P botnets, and (c) 24 hours network traces from a Trans-Pacific backbone line between the United States and Japan as the background network traces (non-P2P \& manually verified P2P).
\begin{table}[!t]
\footnotesize
\captionsetup{font=footnotesize}
\caption{Traces of Legitimate P2P Networks (24 hours)}
\label{table:D1}
\centering
\begin{tabular}{l|c|r|r|c}
\hline
\bfseries Trace & \bfseries \# of hosts & \bfseries \# of flow & \bfseries \# of dstIP & \bfseries Size\\
\hline
  eMule & 16 & 4,181,845 & 725,367 & 42.1G \\
  \hline
  FrostWire & 16 & 4,479,969 & 922,000 & 11.9G \\
  \hline
  uTorrent & 14 & 10,774,924 & 2,326,626 & 57.1G \\
  \hline
  Vuze & 14 & 7,577,039 & 1,208,372 & 20.3G \\
\hline
\end{tabular}
\end{table}

\begin{table}[!t]
\footnotesize
\captionsetup{font=footnotesize}
\caption{Traces of P2P Botnets (24 hours)}
\label{table:D2}
\centering
\begin{tabular}{l|c|r|r|c}
\hline
\bfseries Trace & \bfseries \# of bots & \bfseries \# of flow & \bfseries \# of dstIP & \bfseries Size\\
\hline
  Storm & 13 & 8,603,399 & 145,967 & 5.1G \\
  \hline
  Waledac & 3 & 1,109,508 & 29,972 & 1.1G \\
  \hline
  Sality & 5 & 5,599,440 & 177,594 & 1.5G \\
  \hline
  Kelihos & 8 & 122,182 & 944 & 343.9M \\
  \hline
  ZeroAccess & 8 & 709,299 & 277 & 75.2M \\
\hline
\end{tabular}
\end{table}

\begin{table}[!t]
\footnotesize
\captionsetup{font=footnotesize}
\caption{Traces of Background Network}
\label{table:D3}
\centering
\begin{tabular}{c|c|c|c|c}
\hline
\bfseries Date & \bfseries Dur & \bfseries \# of hosts & \bfseries \# of flow & \bfseries Size\\
\hline
  2014/12/10 & 24 hours & 48,607,304 & 407,523,221 & 788.7G \\
\hline
\end{tabular}
\end{table}

{\bf Legitimate P2P network traces ($D_{p2p}$): }
Our legitimate P2P network traces $D_{p2p}$ were obtained from the University of Georgia \cite{rahbarinia2014peerrush}, which collected the network traces of 4 popular P2P applications for several weeks. We obtained the network traces of 16 eMule hosts, 16 FrostWire hosts, 14 uTorrent hosts and 14 Vuze hosts by randomly selecting a set of continuous 24 hours network traces of each host (as shown in Table~\ref{table:D1}).

{\bf P2P botnets network traces ($D_{bot}$): }
Part of our botnets network traces were from the University of Georgia \cite{rahbarinia2014peerrush}, containing 24 hours network traces of 13 Storm hosts and 3 Waledac hosts. We also collected 24 hours network traces of another three P2P botnets, Sality, Kelihos and ZeroAccess. These network traces were all collected from the hosts manually infected by the binary samples of Kelihos, ZeroAccess, and Sality obtained from \cite{malsamples}. Our data collection was operated in a controlled environment, where all malicious activities were blocked. The same data collection settings were used in several previous works \cite{yan2013peerclean, rahbarinia2014peerrush, zhang2014building}. We collected the network traces of 8 Kelihos bots, 8 ZeroAccess bots and 5 Sality bots (as shown in Table~\ref{table:D2}).

{\bf Background network traces ($D^{b}_{non}$ and $D^{b}_{p2p}$): }
We used a dataset from the MAWI Working Group Traffic Archive \cite{mawi} as our background network traces, containing 24 hours anonymized and payload-free network traces at the transit link of WIDE (150Mbps) to the upstream ISP on 2014/12/10 (as shown in Table~\ref{table:D3}). The dataset contains approximate 407,523,221 flows and 48,607,304 unique IPs. 79.3\% flows are TCP flows and the rest are UDP flows.

We investigated the background network traces, and made our best effort to separate the P2P traffic ($D^{b}_{p2p}$) from the non-P2P traffic ($D^{b}_{p2p}$). Since the WIDE dataset was anonymized and payload-free, it prevented us from using payload analysis to thoroughly check if P2P traffic, especially P2P Botnet traffic existing there. Instead, we used port analysis to manually detect P2P traffic within the background dataset. This is based on the simple concept that many P2P applications have default ports on which they function (see \cite{karagiannis2004transport} for a list of default network ports of popular P2P applications). We manually examined all the network flows of each host in the background network traces. If a host involved in more than five flows using any of the default P2P port values in either source port or destination port, we considered the host as a P2P host. After this procedure, we identified 667 P2P hosts.

One thing worth to be noticed is that despite the whole background network traces lasting for 24 hours, not all these P2P hosts were active for the entire 24 hours. P2P hosts that did not have enough active time, may not produce sufficient network flows for our system to work (as discussed in Section~\ref{sec:sec5_2}). To ensure a fair and rigorous evaluation, we estimated the active time of each P2P host. We divided the 24 hours background network traces into 96 15-minute blocks. If a P2P host had any network flow fell in a block, we considered it was active in that block. We used the number of blocks where a P2P host was active to estimate the active time of each P2P host. Table~\ref{table:P2P_Act} reflects the active time distribution of these P2P hosts. As shown in Table~\ref{table:P2P_Act}, even though there were 667 P2P hosts in total, only 4 of them had been active for the entire 24 hours and 26 of them had been active for no less than 5 hours.

We used ARGUS \cite{argus} to process and cluster network traces into the 5-tuple format tcp/udp flow.
\begin{table}[!t]
\footnotesize
\captionsetup{font=footnotesize}
\caption{Active Time of P2P hosts within the Background Network Trace ($P_{i}$ is the set of P2P hosts have no less than $i \times 15$ minutes active time.)}
\label{table:P2P_Act}
\centering
\begin{tabular}{c|c||c|c||c|c}
\hline
 -  & \bfseries \# of hosts & -  & \bfseries \# of hosts & -  & \bfseries \# of hosts\\
\hline
  $P_{1}$ & 667 & $P_{8}$ & 66 & $P_{32}$ & 21\\
  \hline
  $P_{2}$ & 325 & $P_{14}$ & 38 & $P_{48}$ & 13\\
  \hline
  $P_{4}$ & 180 & $P_{20}$ & 26 & $P_{96}$ & 4\\
  \hline
\end{tabular}
\end{table}

\subsubsection{Experimental Dataset Generation}
\label{sec:sec5_1_3}
As illustrated in Fig.~\ref{fig:network1}, we consider a scenario that an organization has a set of internal hosts communicating with a set of external hosts (outside of the organization), and our system is deployed at the boundary of the organization. Since our original datasets did not maintain a internal-external network structure while collecting them, we generated synthetic experimental datasets by mixing network traces from the original datasets. We considered a case that contains 10,000 internal hosts. For each synthetic experimental dataset, the 667 P2P hosts in $D^{b}_{p2p}$ were considered as the internal hosts. Another 9,333 internal hosts were sampled from $D^{b}_{non}$, where the traffic of 37 randomly selected hosts were mixed with the traffic of 37 P2P bots in $D_{bot}$, and the traffic of another 60 randomly selected hosts were mixed with the traffic of 60 P2P hosts in $D_{p2p}$. To make the experimental evaluation as unbiased and challenging as possible, we propose to sample the internal hosts and generate the synthetic experimental datasets under the following two criterions.
\begin{table}[!t]
\footnotesize
\captionsetup{font=footnotesize}
\caption{Summaries of Experimental Datasets (EDs)}
\label{table:ED}
\centering
\begin{tabular}{l||r}
\hline
\bfseries Descriptions & Values\\
\hline
the \# of EDs & 100\\
\hline
the \# of bots ($D_{bot}$) in each ED & 37\\
\hline
the \# of legitimate P2P hosts ($D_{p2p}$) in each ED & 60\\
\hline
the \# of P2P hosts ($D^{b}_{p2p}$) in each ED & 667\\
\hline
the \# of internal hosts in each ED & 10,000\\
\hline
the AVG \# of external hosts in each ED & 8,642,618\\
\hline
the AVG \# of flow in each ED & 97,640,210\\
\hline
the duration of each ED & 24 hr\\
\hline
\end{tabular}
\end{table}

{\bf Maintain a bipartite network structure.}
Our system aims to deploy at a network boundary (e.g., firewall, gateway, etc.), where the network forms a bipartite structure, and only network flow within the connections between internal hosts and external hosts could be captured. Then, the network in each experimental dataset should maintain a bipartite network structure, where any pair of internal hosts should not have any communications to each other.

{\bf Keep the connectedness of mutual contacts graph.}
The easiest way to obtain a list of background hosts is to sample the hosts randomly from $D^{b}_{non}$, with the respect of bipartite structure. However, since $D^{b}_{non}$ contains an extremely large number of hosts, simply sampling hosts randomly will result in that most of the sampled background hosts do not have a mutual contact with the other background hosts, which is much easier for our system to identify botnet communities. Because less number of mutual contacts among legitimate hosts means more disconnected legitimate communities in the corresponding MCG, which happens to be in favor of Louvain method to detect strongly connected botnet communities. Therefore, we need to sample a list of internal hosts in a way that every internal host should have at least one mutual contact with at least one another internal host.

To follow the criterions described above without making our evaluation tasks any easier, we propose the following synthetic experimental dataset generation procedure:

$\bullet$ Use a two-coloring approach to sample the network traces
%of 9,333 hosts
from $D^{b}_{non}$ without jeopardize the bipartite network structure and the connectedness of mutual contacts graph: (a) initialize two counters, $C_{black}$ and $C_{white}$, to count the number of hosts colored in black and white respectively; (b) coloring a random host $h_{i}$ as black, and $C_{black}$ plus one; (c) coloring all contacts of $h_{i}$ as white, and increase $C_{white}$ by the number of hosts colored as white in this round; (d) for each new colored host, color its contacts with the opposite color, and adjust the counters repeatedly, until we have $C_{black} \geq 9,333$ and $C_{white} \geq 9,333$; (e) select the colored host set with exactly 9,333 hosts as the internal hosts, the hosts in the other set will be the external hosts; and (f) extract the network traces of the 9,333 internal hosts from $D^{b}_{non}$. Then, it forms a bipartite graph, where each colored host set forms a bipartite component, and each host shares at least one mutual contacts with some other hosts from its own bipartite component.

$\bullet$ To maintain a bipartite network structure of botnets and legitimate P2P hosts, we eliminate all communications among bots in $D_{bot}$, and P2P hosts in $D_{p2p}$ and $D^{b}_{p2p}$.

$\bullet$ To mix $D_{bot}$ and $D_{p2p}$ with $D^{b}_{non}$, each time we randomly select 97 internal hosts out of 9,333 background hosts, map the 97 hosts IPs to 37 bots IPs ($D_{bot}$) and 60 legitimate P2P hosts IPs ($D_{p2p}$), and merge the corresponding network traces.

To evaluate our system, 100 synthetic experimental datasets were generated by running this procedure. Table~\ref{table:ED} illustrates the summaries of the experimental datasets (EDs).

\begin{table*}[!t]
\footnotesize
\captionsetup{font=footnotesize}
\caption{Detection Rate and False Positive Rate For Different $\Theta_{dd}$ ($P_{i}$ is the set of P2P hosts within the background network traces that have no less than $i \times 15$ minutes active time. All the hosts of 4 legitimate P2P applications and 5 P2P botnets have 24 hours active time.)}
\label{table:EvaP2P}
\centering
\begin{tabular}{c|c|c|c|c|c|c|c|c|c|c|c|c}
\hline
 \multirow{2}{*}{$\bm \Theta_{dd}$} & \multicolumn{11}{c|}{\bfseries Detection Rate}& \multirow{2}{*}{\bfseries False Positive Rate} \\
 \cline{2-12} & \bfseries Bot & \bfseries P2P & $\bm P_{1}$ & $\bm P_{2}$ & $ \bm P_{4}$ & $\bm P_{8}$ & $\bm P_{14}$ & $\bm P_{20}$ & $\bm P_{32}$ & $\bm P_{48}$ & $\bm P_{96}$  \\
\hline\hline
  2 & 37/37 & 60/60 & 667/667 & 325/325& 180/180& 66/66& 38/38& 26/26& 21/21& 13/13& 4/4& 1,052/9,236\\
  \hline
  5 & 37/37 & 60/60 & 364/667 & 242/325& 180/180& 66/66& 38/38& 26/26& 21/21& 13/13& 4/4& 110/9,236\\
  \hline
  10 & 37/37 & 60/60 & 156/667 & 133/325& 106/180& 66/66& 38/38& 26/26& 21/21& 13/13& 4/4& 44/9,236\\
  \hline
  30 & 37/37 & 60/60 & 36/667 & 36/325& 36/180& 33/66& 30/38& 26/26& 21/21& 13/13& 4/4& 4/9,236\\
  \hline
  50-180 & 37/37 & 60/60 & 15/667 & 15/325& 15/180& 15/66& 15/38& 15/26& 15/21& 13/13& 4/4& 0/9,236\\
  \hline
  185 & 37/37 & 60/60 & 6/667 & 6/325& 6/180& 6/66& 6/38& 6/26& 6/21& 6/13& 4/4& 0/9,236\\
  \hline
  200 & 29/37 & 60/60 & 4/667 & 4/325& 4/180& 4/66& 4/38& 4/26& 4/21& 4/13& 2/4& 0/9,236\\
  \hline
  500-1,000 & 21/37 & 60/60 & 1/667 & 1/325& 1/180& 1/66& 1/38& 1/26& 1/21& 1/13& 1/4& 0/9,236\\
  \hline
  5,000 & 13/37 & 45/60 & 0/667 & 0/325& 0/180& 0/66& 0/38& 0/26& 0/21& 0/13& 0/4& 0/9,236\\
  \hline
  10,000 & 0/37 & 18/60 & 0/667 & 0/325& 0/180& 0/66& 0/38& 0/26& 0/21& 0/13& 0/4& 0/9,236\\
  \hline
  12,500 & 0/37 & 5/60 & 0/667 & 0/325& 0/180& 0/66& 0/38& 0/26& 0/21& 0/13& 0/4& 0/9,236\\
  \hline
  13,500 & 0/37 & 0/60 & 0/667 & 0/325& 0/180& 0/66& 0/38& 0/26& 0/21& 0/13& 0/4& 0/9,236\\
\hline
\end{tabular}
\end{table*}

\subsection{Evaluation on P2P Network Flow Detection}
\label{sec:sec5_2}
We evaluated the P2P network flow detection with different $\Theta_{dd}$. We applied this component on all 100 EDs, and Table~\ref{table:EvaP2P} shows the average detection rate and false positives with different $\Theta_{dd}$, ranging from 2 to 13,500. If $\Theta_{dd}$ is set too small, non-P2P hosts are likely to be detected as P2P hosts, which results in many false positives. For instance, when $2 \leq \Theta_{dd} \leq 5$, at least 110 non-P2P hosts were falsely identified as P2P hosts. If $\Theta_{dd}$ is set too large, all P2P hosts will be removed, which results in false negatives. For instance, when $\Theta_{dd} = 10,000$, most of the P2P hosts were falsely discarded, and only 18 P2P hosts were detected.

On the other hand, the effectiveness of $\Theta_{dd}$ is also subject to the active time of P2P hosts. Since if a P2P host has less active time, it tends to generate less number of P2P network flows to show enough destination diversity, so that it will not be distinguished from non-P2P network flows by our system. For instance, since all the bots and P2P hosts in $D_{bot}$ and $D_{p2p}$ had 24 hours active time, our system can distinguish them well from the non-P2P network flows. However, not all the P2P hosts in $D^{b}_{p2p}$ were active for the entire 24 hours. As shown in Table~\ref{table:EvaP2P}, when the active time of a P2P host was less than 5 hours (not belonging to $P_{20}$, the set of hosts have no less than $20 \times 15$ minutes active time), it was hard for our system to detect P2P network flows from non-P2P network flows ($\Theta_{dd} < 30$).
Hence, when considering P2P hosts that had no less than 12 hours active time ($P_{48}$), and setting $30 \leq \Theta_{dd} \leq 180$, our system detected all P2P hosts with a small number of false positives ($\leq 4/9,236$), which demonstrated that our P2P network flow detection component is stable and effective over a large range of $\Theta_{dd}$ settings.

\subsection{Evaluation on Community Detection}
\label{sec:sec5_3}
\begin{table}[!t]
\footnotesize
\captionsetup{font=footnotesize}
\caption{Community Detection Results For Different $\Theta_{mcr}$}
\label{table:EvaCom}
\centering
\begin{tabular}{c||c|c|c}
\hline
\bfseries $\Theta_{mcr}$ & \bfseries BSI & \bfseries  BAI & \bfseries BLSI\\
\hline
 [0.00, 0.15) & $1.00 \pm 0.00$ & $0.85 \pm 0.00$ & $1.00 \pm 0.00$\\
  \hline
 [0.15, 0.40) & $1.00 \pm 0.00$ & $0.83 \pm 0.02$ & $1.00 \pm 0.00$\\
  \hline
 [0.40, 1.00) & $1.00 \pm 0.00$ & $\leq 0.62 \pm 0.05$ & $1.00 \pm 0.00$\\
\hline
\end{tabular}
\end{table}

We evaluated the performance of community detection with different $\Theta_{mcr}$. We applied this component on the remaining network flows (100 EDs) of the previous component (with $\Theta_{dd}=30$). For each ED, our system generated a MCG $G_{mc} = (V, E)$ with a pre-defined threshold $\Theta_{mcr}$, where each edge $e_{ij} \in E$ contained a weight $mcr_{ij} \in [0.0, 1.0]$. Afterwards, we applied Louvain method (with default resolution 1.0) on the MCG for community detection. The choice of $\Theta_{mcr}$ would have an influence on the community detection results.

We evaluated the community detection performance in terms of (a) the ability to cluster a pair of bots belonging to the same botnet, (b) the ability to separate a pair of bots coming from different botnets, and (c) the ability to separate bots and legitimate applications. As such, we propose three criterions to evaluate the community detection performance below.

Given a set of bots belonging to $n$ botnets $X=\{X_{1}$, $X_{2}$, $\dots$, $X_{n}\}$ (the ground truth), and the community detection results, $m$ communities $Y=\{Y_{1}$, $Y_{2}$, $\dots$, $Y_{m}\}$, define $Bot$ $Separation$ $Index$ (BSI) and $Bot$ $Aggregation$ $Index$ (BAI) as $\text{BSI}=a/(a+c)$ and $\text{BAI}=a/(a+b)$, where $a$ is the number of pairs of bots that are in the same botnet in $X$, and in the same community in $Y$; $b$ is the number of pairs of bots that are in the same botnet in $X$, and in different communities in $Y$; $c$ is the number of pairs of bots that are in different botnets in $X$, and in the same community in $Y$. BSI denotes the degree of that bots coming from different botnets being separated into different communities. BAI denotes the degree of that bots coming from the same botnet being clustered into the same community. Both BSI and BAI are between 0.0 and 1.0, and the higher the better. ``BSI equals to 1.0'' means all different types of bots are well separated, and ``BAI equals to 1.0'' means all the same types of bots are well clustered.

Given $p$ bots and $q$ legitimate applications, define $Bot$-$Legitimate$ $Separation$ $Index$ (BLSI) as $\text{BLSI}=d/(p \times q)$, where $d$ is the number of pairs of a bot and a legitimate application being separated into different communities via our method. BLSI indicates the ability of our method to separate bots and legitimate applications. BLSI is between 0.0 and 1.0, and the higher the better. ``BLSI equals to 1.0'' means all pairs of one bot and one legitimate application are well separated.

Table~\ref{table:EvaCom} shows the community detection results with different $\Theta_{mcr}$, ranging from 0.0 to 1.0. If $\Theta_{mcr}$ is set too small, there will be more non-zero weight edges, which might result in less but larger communities. On the other hand, if $\Theta_{mcr}$ is set too large, most of the vertices will be isolated, which results in more but smaller communities.
For instance, as $\Theta_{mcr}$ increasing, BSI decreased. When $\Theta_{mcr} \leq 0.4$, BSI was around 0.8 to 0.85, meaning one or more botnets have been split into different communities. It turned out to be our algorithm separates the Storm botnet (13 bots) into two communities, one containing 10 bots and another containing 3 bots. Changing $\Theta_{mcr}$ does not affect BSI and BLSI.
BSI=1.0 means our system separates different types of bots into different communities. BLSI=1.0 means our system separates bots and legitimate P2P applications into different communities. The result demonstrated that our system is very effective and robust in separating bots and legitimate hosts, and separating different types of bots. Since larger $\Theta_{mcr}$ will result in less edges in the MCG, which could reduce the execution time of community detection, we used $\Theta_{mcr}= 0.1$ as our system parameter.

\subsection{Evaluation on Botnet Detection}
\label{sec:sec5_4}
\begin{table}[!t]
\footnotesize
\captionsetup{font=footnotesize}
\centering\caption{Botnet Detection Results For Different $\Theta_{avgddr}$ and $\Theta_{avgmcr}$. (ZeroA.: the detection rate of ZeroAccess; FP: the number of false positives.)}
\label{table:EvaBot}
\centering
\begin{tabular}{c|c|c|c|c|c|c}
\hline
 &  & \multicolumn{5}{c}{$\Theta_{avgddr}$}\\
  \hline
  $\Theta_{avgmcr}$ & - & 0.0 & 0.2 & 0.4 & 0.6 & 0.8\\
  \hline
  \multirow{9}{*}{0.0}& ZeroA. & 100\% & 100\% & 100\% & 100\% & 100\%\\
   & Waledac & 100\% & 100\% & 100\% & 100\% & 100\%\\
   & Storm & 100\% & 100\% & 100\% & 100\% & 100\%\\
   & Kelihos & 100\% & 100\% & 100\% & 100\% & 100\%\\
   & Sality & 100\% & 100\% & 100\% & 100\% & 0\%\\
   & Precision & 28.9\% & 29.1\% & 29.3\% & 38.1\% & 34.8\%\\
   & Recall & 100\% & 100\% & 100\% & 100\% & 86.5\%\\
   & FP & 91 & 90 & 89 & 60 & 60\\
   & F-score & 44.8\% & 45.1\% & 45.4\% & 55.2\% & 49.6\%\\
 \hline
  \multirow{9}{*}{0.05}& ZeroA. & 100\% & 100\% & 100\% & 100\% & 100\%\\
   & Waledac & 100\% & 100\% & 100\% & 100\% & 100\%\\
   & Storm & 100\% & 100\% & 100\% & 100\% & 100\%\\
   & Kelihos & 100\% & 100\% & 100\% & 100\% & 100\%\\
   & Sality & 100\% & 100\% & 100\% & 100\% & 0\%\\
   & Precision & 33.9\% & 34.2\% & 34.9\% & 47.4\% & 43.8\%\\
   & Recall & 100\% & 100\% & 100\% & 100\% & 86.5\%\\
   & FP & 72 & 71 & 69 & 41 & 41\\
   & F-score & 50.7\% & 51\% & 51.7\% & 64.3\% & 58.2\%\\
  \hline
  \multirow{9}{*}{0.1}& ZeroA. & 100\% & 100\% & 100\% & 100\% & 100\%\\
   & Waledac & 100\% & 100\% & 100\% & 100\% & 100\%\\
   & Storm & 100\% & 100\% & 100\% & 100\% & 100\%\\
   & Kelihos & 100\% & 100\% & 100\% & 100\% & 100\%\\
   & Sality & 100\% & 100\% & 100\% & 100\% & 0\%\\
   & Precision & 56.0\% & 56.9\% & 56.9\% & 100\% & 100\%\\
   & Recall & 100\% & 100\% & 81\% & 100\% & 86.5\%\\
   & FP & 29 & 28 & 28 & 0 & 0\\
   & F-score & 71.8\% & 72.5\% & 72.5\% & 100\% & 92.8\%\\
   \hline
  \multirow{9}{*}{0.15-0.35}& ZeroA. & 100\% & 100\% & 100\% & 100\% & 100\%\\
   & Waledac & 100\% & 100\% & 100\% & 100\% & 100\%\\
   & Storm & 100\% & 100\% & 100\% & 100\% & 100\%\\
   & Kelihos & 100\% & 100\% & 100\% & 100\% & 100\%\\
   & Sality & 100\% & 100\% & 100\% & 100\% & 0\%\\
   & Precision & 100\% & 100\% & 100\% & 100\% & 100\%\\
   & Recall & 100\% & 100\% & 100\% & 100\% & 86.5\%\\
   & FP & 0 & 0 & 0 & 0 & 0\\
   & F-score & 100\% & 100\% & 100\% & 100\% & 92.8\%\\
    \hline
  \multirow{9}{*}{0.4}& ZeroA. & 100\% & 100\% & 100\% & 100\% & 100\%\\
   & Waledac & 100\% & 100\% & 100\% & 100\% & 100\%\\
   & Storm & 84.6\% & 84.6\% & 84.6\% & 84.6\% & 76.9\%\\
   & Kelihos & 100\% & 100\% & 100\% & 100\% & 100\%\\
   & Sality & 100\% & 100\% & 100\% & 100\% & 0\%\\
   & Precision & 100\% & 100\% & 100\% & 100\% & 100\%\\
   & Recall & 94.6\% & 94.6\% & 94.6\% & 94.6\% & 78.4\%\\
   & FP & 0 & 0 & 0 & 0 & 0\\
   & F-score & 97.2\% & 97.2\% & 97.2\% & 97.2\% & 87.9\%\\
   \hline
  \multirow{9}{*}{0.6-0.8}& ZeroA. & 100\% & 100\% & 100\% & 100\% & 100\%\\
   & Waledac & 100\% & 100\% &100\% & 100\% & 0\%\\
   & Storm & 0\% & 0\% & 0\% & 0\% & 0\%\\
   & Kelihos & 100\% & 100\% & 100\% & 100\% & 100\%\\
   & Sality & 100\% & 100\% & 100\% & 100\% & 0\%\\
   & Precision & 100\% & 100\% & 100\% & 100\% & 100\%\\
   & Recall & 64.9\% & 64.9\% & 64.9\% & 64.9\% & 43.2\%\\
   & FP & 0 & 0 & 0 & 0 & 0\\
   & F-score & 78.7\% & 78.7\% & 78.7\% & 78.7\% & 60.4\%\\
\hline
\end{tabular}
\end{table}

We evaluated the botnet detection component with different parameter settings. We applied this component on the remaining network flows (100 EDs) of the previous component (with $\Theta_{dd}=30$ and $\Theta_{mcr}=0.1$). We assumed that all the host in the background trace ($D^{b}$ and $D^{b}_{p2p}$) were not malicious, and would be reported as false positives if being detected.

Table~\ref{table:EvaBot} shows the P2P botnet detection results which supports our idea that the AVGDDR of legitimate P2P network flow cluster communities is lower than most of the P2P botnets network flow cluster communities. For instance, the AVGDDR of all (60/60) legitimate P2P network flow cluster communities were higher than 0.6, and the AVGDDR of 32 out of 37 botnets were higher than $0.8$. The other 5 turned out to be 5 Sality bots, which could be detected by AVGMCR. Also, the legitimate P2P network flow clusters have lower AVGMCR than P2P bots (i.e., $\Theta_{avgmcr} \in [0.15, 0.35]$). For most of the botnets (i.e., ZeroAccess, Waledac, Kelihos and Sality), our system is effective (100\% detection rate with zero false positive) and stable over a large range of $\Theta_{avgddr}$ (i.e., $[0.0, 0.6]$) and $\Theta_{avgmcr}$ (i.e., $[0.15, 0.8]$). Storm has a relative small AVGMCR, hence the effective parameters narrowed down to $\Theta_{avgddr} \in [0.0, 0.6]$ and $\Theta_{avgmcr} \in [0.15, 0.35]$.

\begin{table*}[!t]
\footnotesize
\captionsetup{font=footnotesize}
\caption{The number of hosts identified by each component}
\label{table:EvaNum}
\centering
\begin{tabular}{c|c|c|c|c}
\hline
 - & \bfseries Before P2P detection & \bfseries P2P detection & \bfseries Community detection & \bfseries Botnet detection\\
\hline
  \bfseries \# of hosts & 10,000 & 97 & 97 & 37\\
\hline
\end{tabular}
\end{table*}

\begin{table*}[!t]
\footnotesize
\captionsetup{font=footnotesize}
\caption{Enhanced PeerHunter Execution Time}
\label{table:EvaTime}
\centering
\begin{tabular}{c|c|c|c|c|c}
\hline
 - & \bfseries P2P Network Flow Detection & \bfseries MCG Extraction & \bfseries Community Detection & \bfseries Bot Detection & \bfseries Total \\
\hline
 \bfseries Processing Time  & 15 minutes & 5 minutes & 5 seconds & 10 seconds & 20 minutes\\
\hline
\end{tabular}
\end{table*}

\subsection{Evaluation on Enhanced PeerHunter}
\label{sec:sec5_5}
\subsubsection{Analyzing the System Effectiveness}
\label{sec:sec5_5_1}
We applied Enhanced PeerHunter on 100 EDs, with $\Theta_{dd}$=$30$, $\Theta_{mcr}$=$0.1$, $\Theta_{avgddr}$=$0.6$ and $\Theta_{avgmcr}$=$0.15$, and all the results were averaged over 100 EDs. Using $\Theta_{avgddr}$=$0.6$ and $\Theta_{avgmcr}$=$0.15$ was based on our empirical study (shown in Table~\ref{table:EvaBot}). As illustrated in Table~\ref{table:EvaNum}, our system identified all 97 P2P hosts from 10,000 hosts, and detected all 37 bots from those 97 P2P hosts, with zero false positive, which demonstrated that Enhanced PeerHunter is effective and accurate in detecting P2P botnets.

\subsubsection{Analyzing the System Scalability}
\label{sec:sec5_5_2}
\begin{figure}[tb]
\captionsetup{font=footnotesize}
    \centering
        \includegraphics[width=0.4\textwidth]{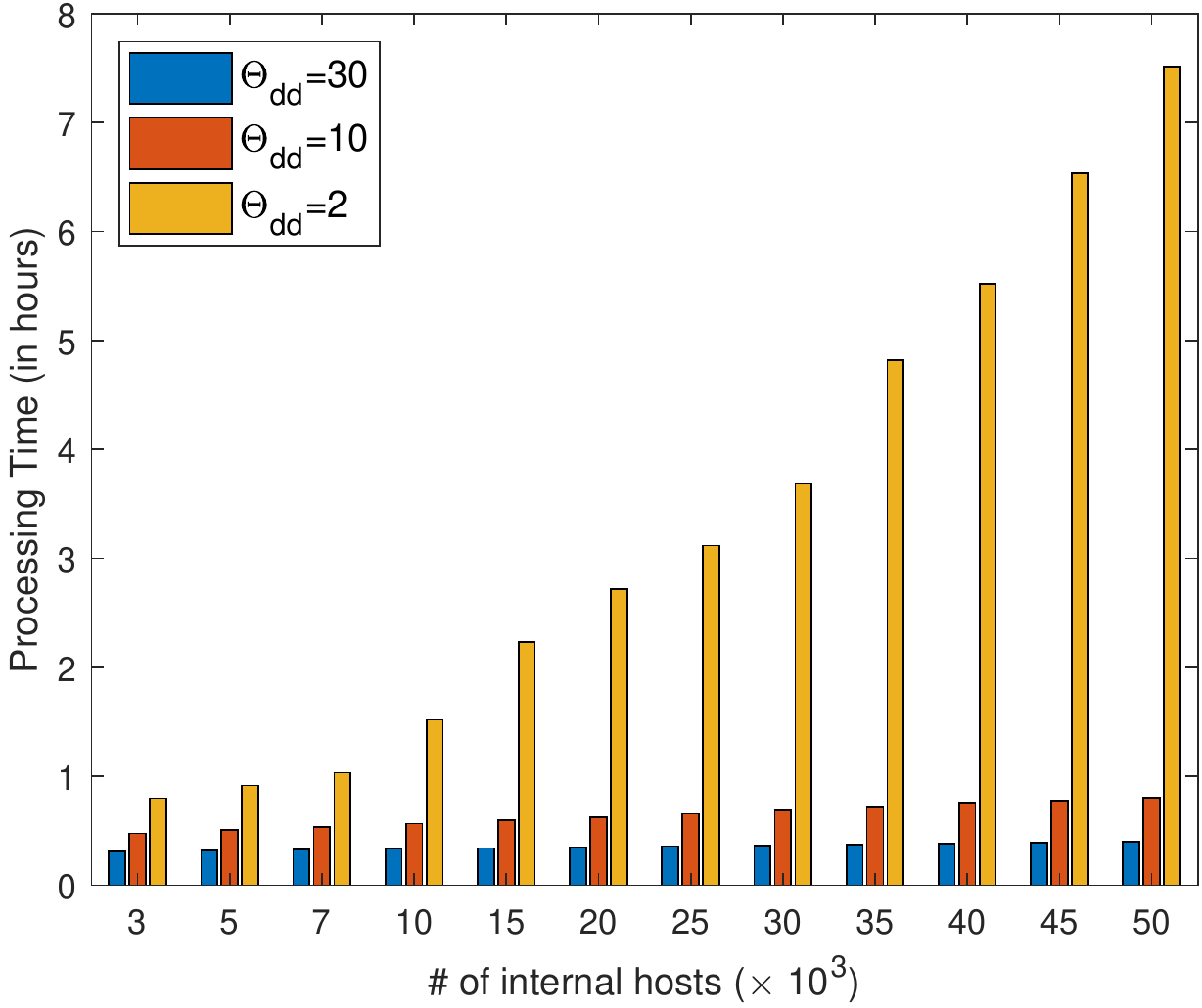}
        \caption{Processing time with different data size and $\Theta_{dd}$.}
        \label{fig:time}
\end{figure}

The system scalability is to evaluate the practicality of our systems to deal with the real world big data. First, we applied Enhanced PeerHunter on 100 EDs of 10,000 internal hosts to analyze the processing time of each component. 
Our system has a scalable design based on efficient detection algorithm and distributed/parallelized computation. 
As shown in Table~\ref{table:EvaNum}, community detection and botnet detection had negligible processing time compared with P2P network flow detection and MCG extraction, since our first two steps (i.e., P2P network flow detection and MCG extraction) were designed to reduce a huge amount of the hosts subject to analysis (i.e., 99.03\% in our experiments).
The P2P network flow detection component has linear time complexity, since it scans all the input flows only once to get the flow clusters and further detect P2P flow clusters. However, since it is the very first component to process the input data (data could be large), it still costs the highest processing time (i.e., 15 minutes). 
To accommodate the growth of a real-world input data, we designed and implemented the P2P network flow detection component using a MapReduce framework, which could be deployed in distributed fashion on scalable cloud computing platforms (e.g., amazon EC2). 
The MCG extraction component requires pairwise comparison to calculate edges weights.
Let $n$ be the number of P2P network flow clusters subject to analysis and $m$ be the maximum number of distinct contacts of a P2P network flow cluster.
We implemented the comparison between each pair of hosts parallelly to handle the growth of $n$. If we denote $k$ as the number of threads running parallelly, the time complexity of MCG extraction is $O(\frac{n^{2}m}{k})$. 
For a given ISP network, $m$ grows over time. Since our system uses a fixed time window (24 hours), for a given ISP network, $m$ tends to be stable and would not cause a scalability issue. Besides, since the percentage of P2P hosts of an ISP network is relatively small (i.e., $3\%$ \cite{zhang2014building}), an ISP network usually has less than 65,536 (/16 subnet) hosts, and most P2P hosts generate less than 150 P2P network flow clusters (our empirical study), $n$ would be negligible compared with $m$. Moreover, since the waiting stage bots always act stealthily and only make necessary communications, $m$ also will not be large. 
We also tested our system using different sizes (i.e., different number of internal hosts) of EDs. For each size, we generated 10 EDs, and recorded the average processing time of our system with different $\Theta_{dd}$. As shown in Fig.~\ref{fig:time}, compared with the size of datasets, $\Theta_{dd}$ has more influence on the system scalability. Because in our P2P network flow detection component, $\Theta_{dd}$ has an impact on $n$ (the number of P2P network flow clusters subject to analysis), and larger $\Theta_{dd}$ leads to smaller $n$, thus less processing time. For instance, when $\Theta_{dd}=10$ or $30$, the increase of processing time, caused by increasing the size of data, was much less than when $\Theta_{dd}=2$.
Therefore, our system is very scalable on different sizes of data with an appropriate $\Theta_{dd}$ (e.g., 10 or 30). Also, by tuning $\Theta_{dd}$, our system has the potential to deal with different size of datasets in a reasonable time. To summarize, Enhanced PeerHunter is scalable to handle the real world network data.

\subsubsection{Analyzing the Effectiveness of System Parameters}
\label{sec:sec5_5_3}
Although we had analyzed the effectiveness of $\Theta_{dd}$, $\Theta_{avgddr}$ and $\Theta_{avgmcr}$ within the corresponding components, the effectiveness of combinations among different values of $\Theta_{dd}$, $\Theta_{avgddr}$ and $\Theta_{avgmcr}$ has not been studied.
As shown in Fig.~\ref{fig:Para_26}, we used precision, recall and false positives to evaluate the effectiveness of different parameter combinations. As discussed in Section~\ref{sec:sec5_2}, $\Theta_{dd}$ is used to detect P2P network flow clusters. Larger $\Theta_{dd}$ tends to result in more false negatives (lower recall), 
and smaller $\Theta_{dd}$ tends to result in more false positives (lower precision). 
For instance, changing $\Theta_{dd}$ from 30 or 50 to 10 resulted in 47 or 42 more false positives ($\Theta_{avgddr}=0.15$) as shown in Fig.~\ref{fig:Para2_FP} and Fig.~\ref{fig:Para6_FP}, respectively.
When $\Theta_{dd} \in \{30, 50\}$, $\Theta_{avgddr} \in [0.15, 0.35]$ and $\Theta_{avgddr} \in [0.2, 0.6]$, our system yielded 100\% detection rate with zero false positive. Even when $\Theta_{dd}=10$, our system can still work effectively with $\Theta_{avgddr} \in [0.25, 0.35]$ and $\Theta_{avgddr} \in [0.2, 0.6]$. This demonstrated our system can work effectively over several different parameter combinations.

\subsubsection{Analyzing the ``True'' False Positives when $\Theta_{dd}=10$}
\label{sec:sec5_5_4}
\begin{figure*}[tb]
\captionsetup{font=footnotesize}
        \centering
        \begin{subfigure}[b]{0.32\textwidth}
                \includegraphics[width=\textwidth]{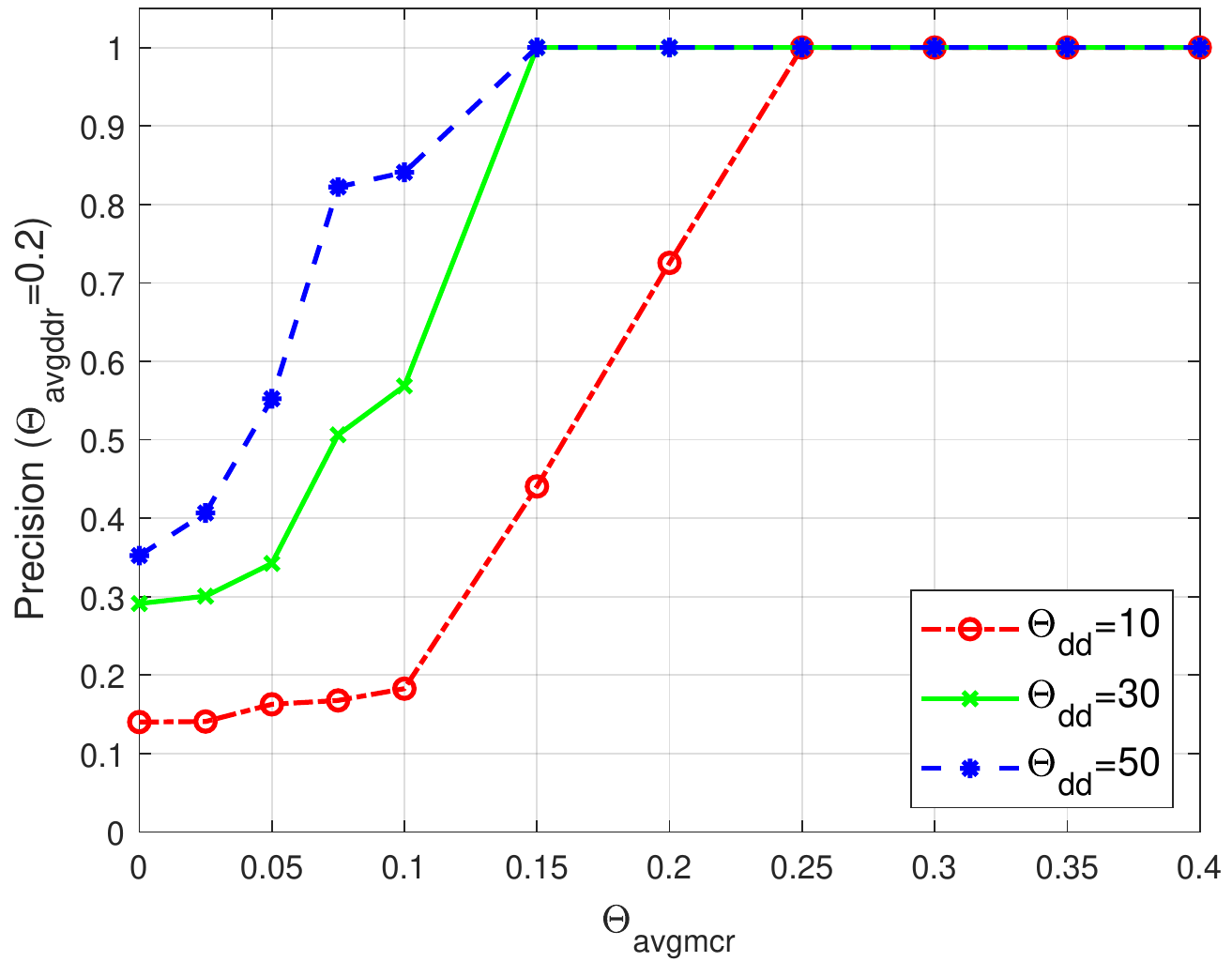}
                \caption{}
                \label{fig:Para2_Precision}
        \end{subfigure}%
        ~ %add desired spacing between images, e. g. ~, \quad, \qquad, \hfill etc.
        \begin{subfigure}[b]{0.32\textwidth}
                \includegraphics[width=\textwidth]{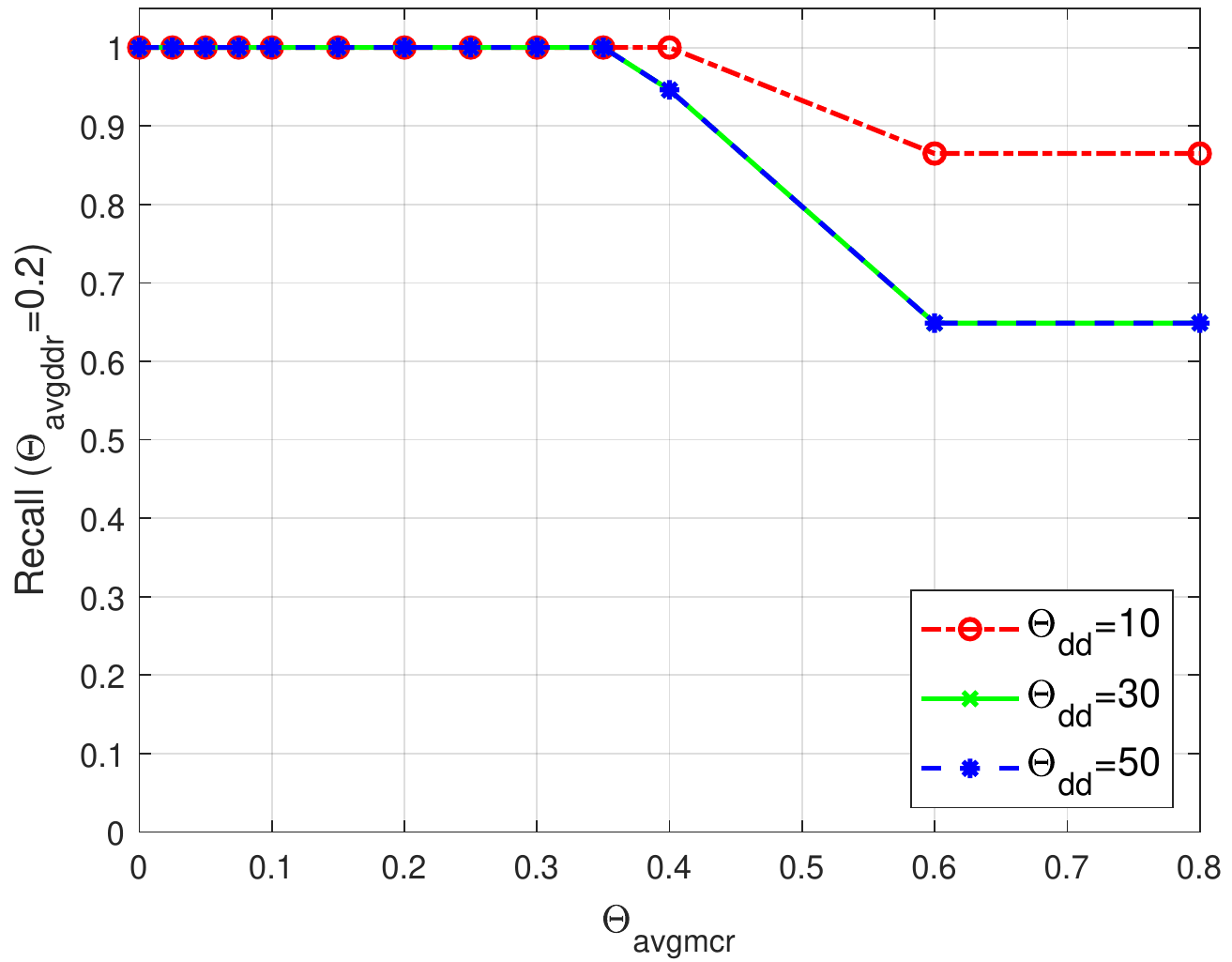}
                \caption{}
                \label{fig:Para2_Recall}
        \end{subfigure}
        ~ %add desired spacing between images, e. g. ~, \quad, \qquad, \hfill etc.
        \begin{subfigure}[b]{0.32\textwidth}
                \includegraphics[width=\textwidth]{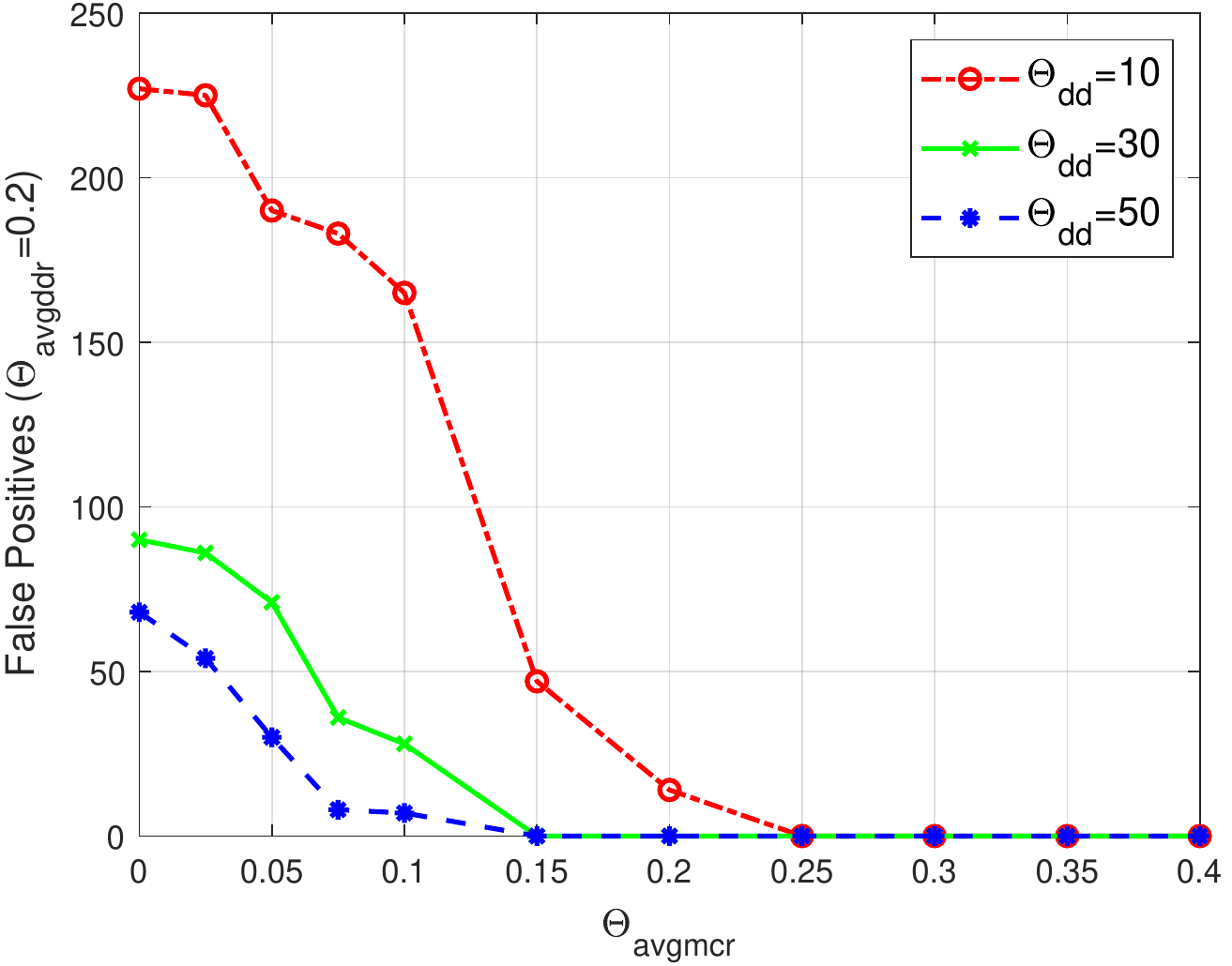}
                \caption{}
                \label{fig:Para2_FP}
        \end{subfigure}
        ~ %add desired spacing between images, e. g. ~, \quad, \qquad, \hfill etc.
        \begin{subfigure}[b]{0.32\textwidth}
                \includegraphics[width=\textwidth]{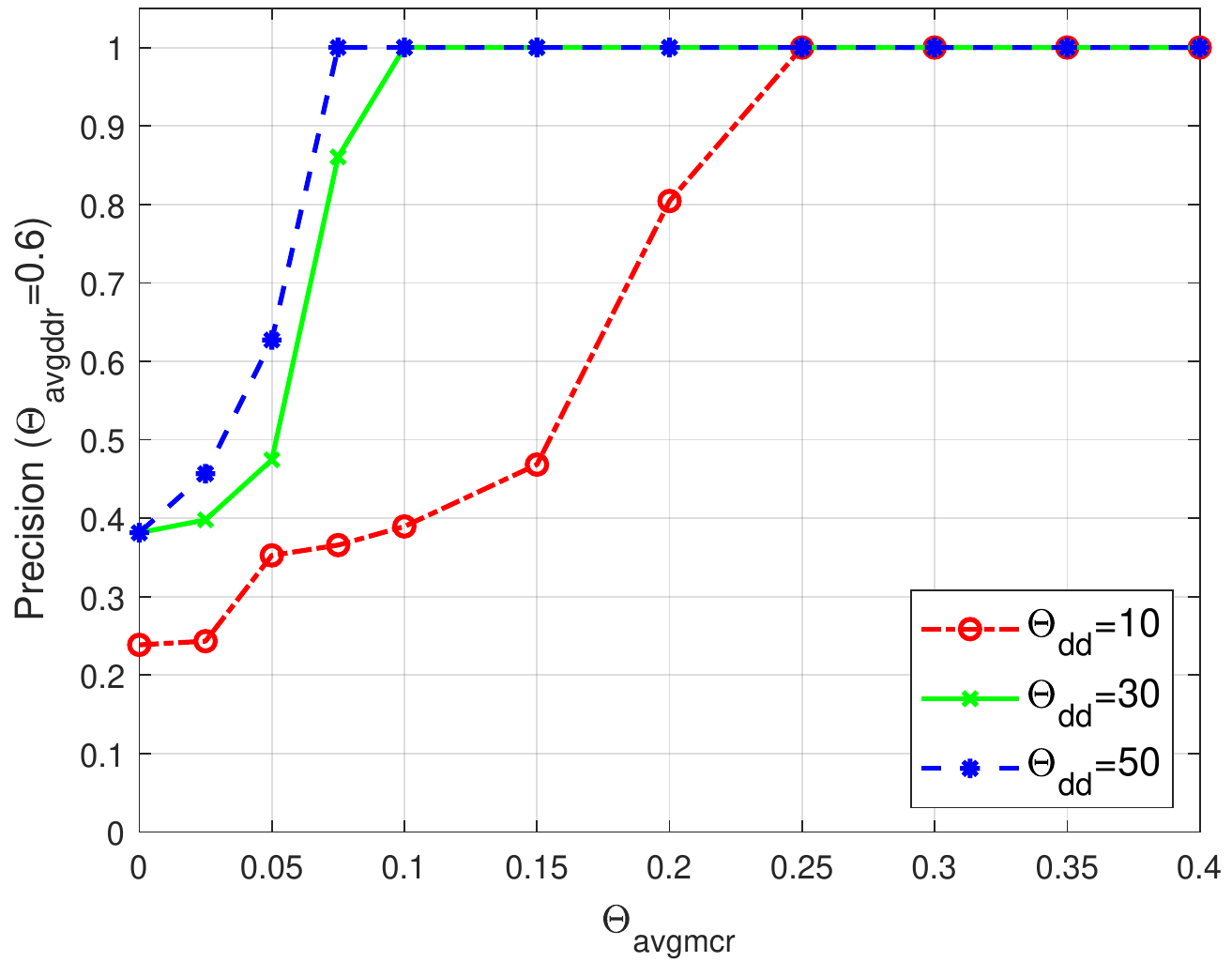}
                \caption{}
                \label{fig:Para6_Precision}
        \end{subfigure}%
        ~ %add desired spacing between images, e. g. ~, \quad, \qquad, \hfill etc.
        \begin{subfigure}[b]{0.32\textwidth}
                \includegraphics[width=\textwidth]{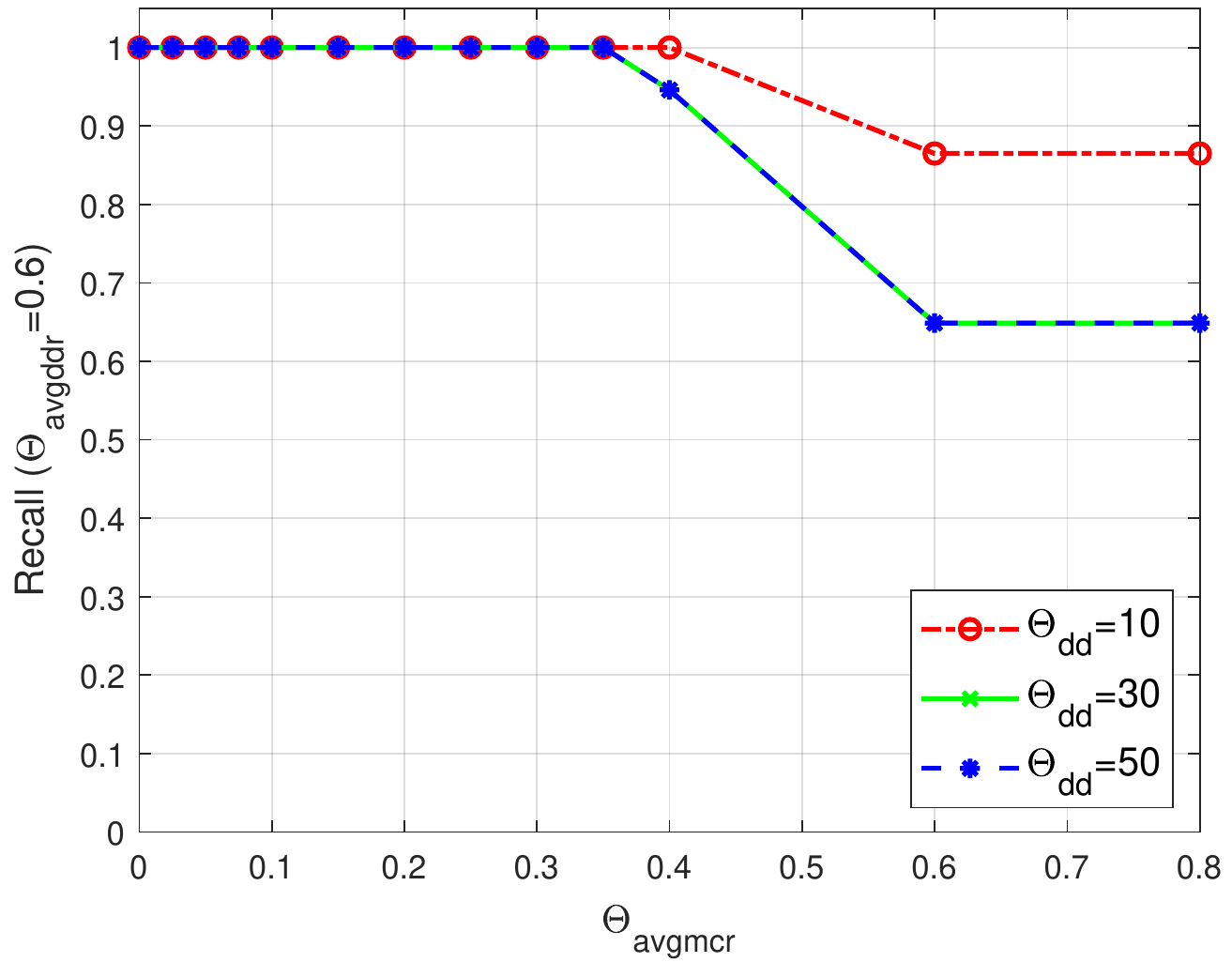}
                \caption{}
                \label{fig:Para6_Recall}
        \end{subfigure}
        ~ %add desired spacing between images, e. g. ~, \quad, \qquad, \hfill etc.
        \begin{subfigure}[b]{0.32\textwidth}
                \includegraphics[width=\textwidth]{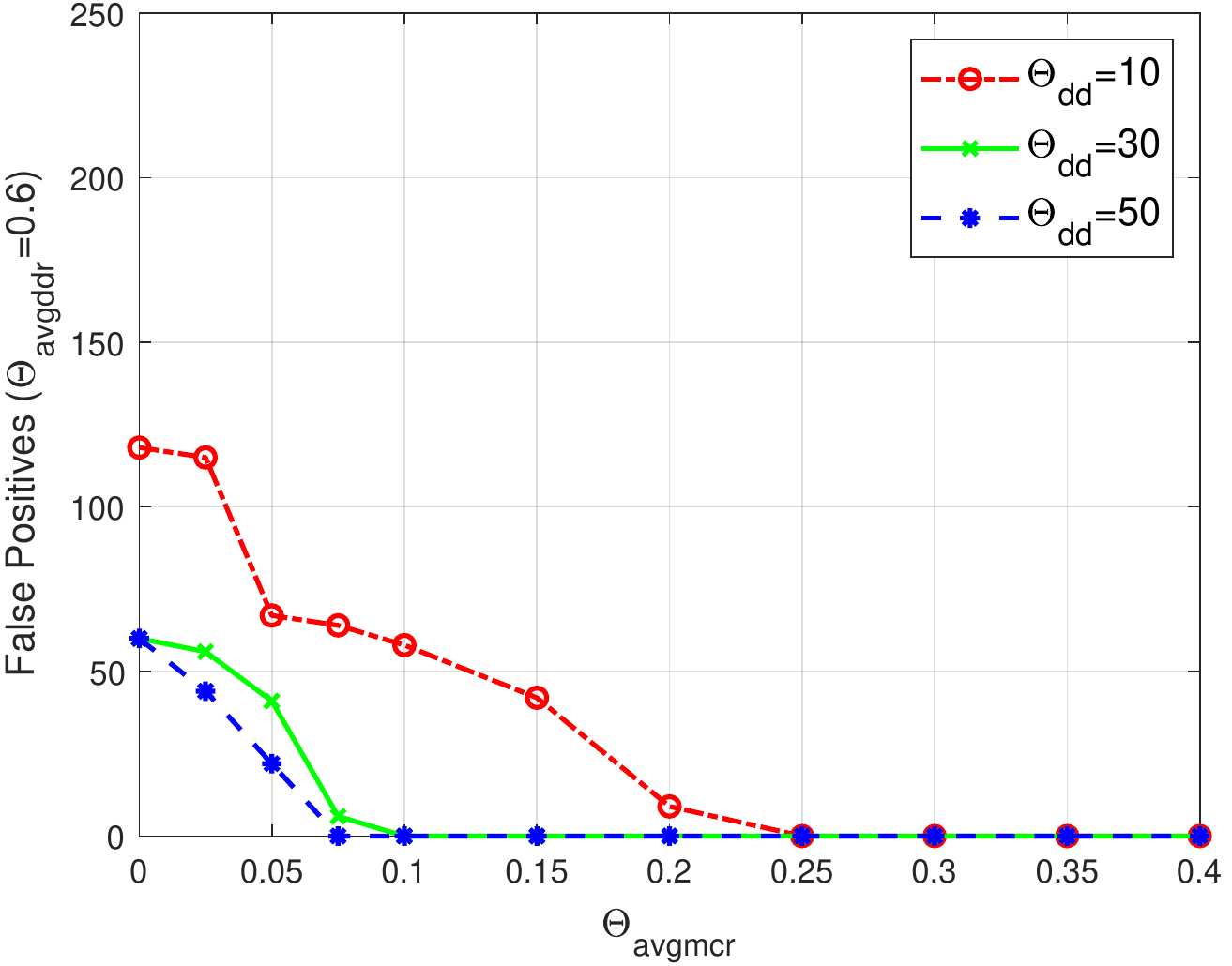}
                \caption{}
                \label{fig:Para6_FP}
        \end{subfigure}
        \caption{Precision, recall and false positives given different $\Theta_{dd}$, $\Theta_{avgddr}$ and $\Theta_{avgmcr}$ ($\Theta_{mcr}=0.05$).}
        \label{fig:Para_26}
\end{figure*}

In this section, we discuss about some interesting findings about the false positives resulted from setting $\Theta_{dd}=10$. As discussed in Section~\ref{sec:sec3_3_2}, $\Theta_{avgddr}$ is used to capture the ``P2P behavior'' of network flows, and $\Theta_{avgmcr}$ is used to capture the ``botnet behavior'' of network flows. Hence, if we use a larger $\Theta_{avgddr}$ (i.e., 0.6) and a smaller $\Theta_{avgmcr}$ (i.e., 0.0), most of the false positives should be legitimate P2P host. For instance, in Fig.~\ref{fig:Para6_FP}, when $\Theta_{dd}=10$, $\Theta_{avgddr}=0.6$ and $\Theta_{avgmcr}=0.0$, 115 out of 118 false positives were P2P hosts (60 from $D_{p2p}$ and 55 from $D^{b}_{p2p}$).
On the other hand, we assume that if we use a smaller $\Theta_{avgddr}$ (i.e., 0.2) and a larger $\Theta_{avgmcr}$ (i.e., 0.15), some of the false positives might come from the other types of botnets. As shown in Fig.~\ref{fig:Para2_FP}, when $\Theta_{dd}=10$, $\Theta_{avgddr}=0.2$ and $\Theta_{avgmcr}=0.15$, 9 out of 47 false positives were not our known legitimate P2P hosts. We investigated these false positives, with their anonymized and payload-free network traces. It turned out that, 4 out of the 9 false positives (i.e., ``180.217.2.181'', ``180.217.2.246 '', ``180.217.2.248'' and ``180.217.2.177'') were listed in the Barracuda Reputation Block List (BRBL) \cite{BRBL}, a highly accurate list of the IP addresses known to send spam.
Hence, we are convinced that those false positives were infected with virus or botnets. These interesting ``true'' false positives findings demonstrated that our system has the potential to detect other unknown botnets.

\subsection{Mimicking Legitimate P2P Application Attacks (MMKL)}
\label{sec:sec5_6}
Our work is focusing on detecting P2P botnets from legitimate P2P applications. 
If the adversaries (e.g., botmasters) know our techniques in advance, they might attempt to evade our system via instructing P2P bots to mimic the behavior of legitimate P2P applications. Inspired by \cite{zhang2014building}, in this section, we propose two evasion attacks. 
All the parameters used in experiments of this Section were the same as in Section\ref{sec:sec5_5}.

\begin{table}[!t]
\footnotesize
\captionsetup{font=footnotesize}
\centering\caption{Comparison of the community detection results between PeerHunter \cite{zhuang2017peerhunter} and Enhanced PeerHunter under PMMKL.}
\label{table:Eva_ATT1_CD}
\centering
\begin{tabular}{c|c|c|c|c}
\hline
\multirow{2}{*}{-} &
\multicolumn{2}{c|}{PeerHunter \cite{zhuang2017peerhunter}} &
\multicolumn{2}{c}{Enhanced PeerHunter}\\
\cline{2-5}  & No Attack & PMMKL & No Attack & PMMKL \\
\hline
BSI & $1.00 \pm 0.00$ & $0.73 \pm 0.02$ & $1.00 \pm 0.00$ & $1.00 \pm 0.00$ \\
\hline
BAI & $1.00 \pm 0.00$ & $0.81 \pm 0.01$ & $0.85 \pm 0.00$ & $0.85 \pm 0.00$ \\
\hline
BLSI & $1.00 \pm 0.00$ & $0.78 \pm 0.01$ & $1.00 \pm 0.00$ & $1.00 \pm 0.00$ \\
\hline
\end{tabular}
\end{table}

\begin{table*}[!t]
\footnotesize
\captionsetup{font=footnotesize}
\centering\caption{Comparison of the botnet detection results under no attack and PMMKL attack. (* detection rate)}
\label{table:Eva_ATT1_BD}
\centering
\begin{tabular}{c|c|c|c|c|c|c|c|c}
\hline
\multirow{2}{*}{-} &
\multicolumn{2}{c|}{PeerHunter \cite{zhuang2017peerhunter}} &
\multicolumn{2}{c|}{Enhanced PeerHunter} &
\multicolumn{2}{c|}{Zhang \textit{et al.} \cite{zhang2014building} ($\Theta_{bot}=0.6$)} &
\multicolumn{2}{c}{Zhang \textit{et al.} \cite{zhang2014building} ($\Theta_{bot}=0.8$)} \\
\cline{2-9} & No Attack & PMMKL & No Attack & PMMKL & No Attack & PMMKL & No Attack & PMMKL\\
\hline
ZeroAccess* & 100\% & 0\% &  100\% & 100\% &  100\% & 82.5\% &  100\% & 90\%\\
\hline
Waledac* & 100\% & 0\% &  100\% & 100\% &  100\% & 0\% &  100\% & 60\%\\
\hline
Storm* & 100\% & 37.5\% &  100\% & 100\% &  97.8\% & 61.5\% &  100\% & 95.4\%\\
\hline
Kelihos* & 100\% & 0\% &  100\% & 100\% &  85.5\% & 45\% &  85.5\% & 77.5\%\\
\hline
Sality* & 100\% & 79.2\% &  100\% & 100\% &  89.6\% & 80\% &  96.8\% & 88\%\\
\hline
Precision & 100\% & 99.1\% &  100\% & 100\% &  100\% & 100\% &  60.8\% & 62.7\%\\
\hline
Recall & 100\% & 23.9\% &  100\% & 100\% &  94.7\% & 60\% &  96.4\% & 86.5\%\\
\hline
FP & 0/9,963 & 39/9,926 &  0/9,963 & 0/9,926 &  0/9,963 & 0/9,926 &  23/9,963 & 19/9,926\\
\hline
F-score & 100\% & 38.5\% &  100\% & 100\% &  97.3\% & 75\% &  74.6\% & 72.7\%\\
\hline
\end{tabular}
\end{table*}

\subsubsection{Passive MMKL (PMMKL)}
\label{sec:sec5_6_1}
In this attack, the botmaster can instruct the bots to passively generate network traffic together with legitimate P2P applications running on the same machine at the same time. As such, the botnet traffic will be overlapped with the legitimate P2P traffic. Since during most of the time, P2P botnets will be acting stealthily, the legitimate P2P traffic will dominate the host level behavior. Hence, the attack could effectively evade the host level group behavior based methods \cite{yan2013peerclean, zhuang2017peerhunter}.
Also, the attack does not require the botnets to generate more or new types of network flows, and just need to monitor the legitimate P2P application activities, which can evade certain anomaly-based methods. Since our detection algorithm is based on network-flow level mutual contacts graph, which could differentiate the network flows coming from different P2P applications, it is capable of detecting P2P bots while the bots traffic and the legitimate P2P traffic are overlapped on the same host.

To simulate this attack on each ED, we randomly selected 37 hosts out of the 60 legitimate P2P application hosts, and randomly mapped their IPs to 37 bots' IPs. By doing this, the traffic of each bot were overlapped with the traffic of one legitimate P2P host.
And we made a comparison between Enhanced PeerHunter and PeerHunter \cite{zhuang2017peerhunter} under this attack, where PeerHunter \cite{zhuang2017peerhunter} was using one of its best parameter setting $\Theta_{dd}$=$50$, $\Theta_{mcr}$=$0.05$, $\Theta_{avgddr}$=$0.06$ and $\Theta_{avgmcr}$=$0.2$.
As shown in Table~\ref{table:Eva_ATT1_CD}, all three community detection indices (i.e., BSI, BAI and BLSI) decreased around 20\% while running PeerHunter under this attack. However, PMMKL had no effects on Enhanced PeerHunter's community detection performance.
As shown in Table~\ref{table:Eva_ATT1_BD}, PMMKL completely failed PeerHunter in detecting ZeroAccess, Waledac and Kelihos, and dramatically reduced the detection rate of Storm and Sality. On the contrary, PMMKL had no affects on Enhanced PeerHunter's P2P botnet detection performance.

To summarize, compared with our previous work, Enhanced PeerHunter can detect P2P botnets effectively even if bots are running on the same host as legitimate P2P applications.

\subsubsection{Active MMKL (AMMKL)}
\label{sec:sec5_6_2}
In this attack, the botmaster can instruct the bots to mimic the behaviors of legitimate P2P applications actively. For instance, each bot can actively communicate with an extra set of randomly selected peers to decrease the rate of mutual contacts between a pair of bots. Compared with PMMKL, in AMMKL, bots do not need to monitor and wait until some legitimate P2P application running to work. However, communicating with much more extra but unnecessary peers will lead the botnets to act less stealthy and less efficient, and enable certain anomaly-based methods (e.g., high volumes of network traffic) to detect them.

\begin{figure*}[tb]
\captionsetup{font=footnotesize}
        \centering
        \begin{subfigure}[b]{0.32\textwidth}
                \includegraphics[width=\textwidth]{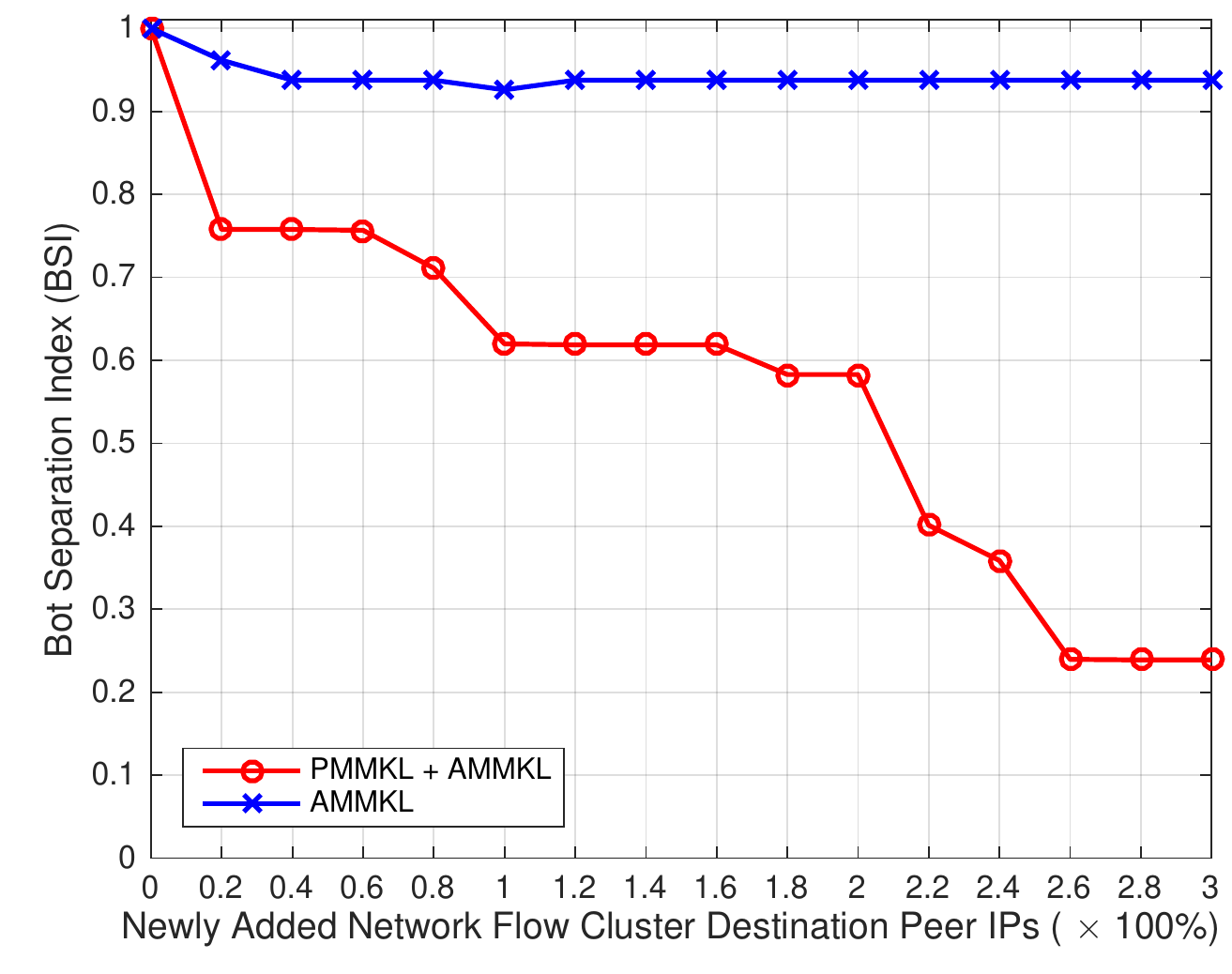}
                \caption{}
                \label{fig:ATT2_BSI}
        \end{subfigure}%
        ~ %add desired spacing between images, e. g. ~, \quad, \qquad, \hfill etc.
        \begin{subfigure}[b]{0.32\textwidth}
                \includegraphics[width=\textwidth]{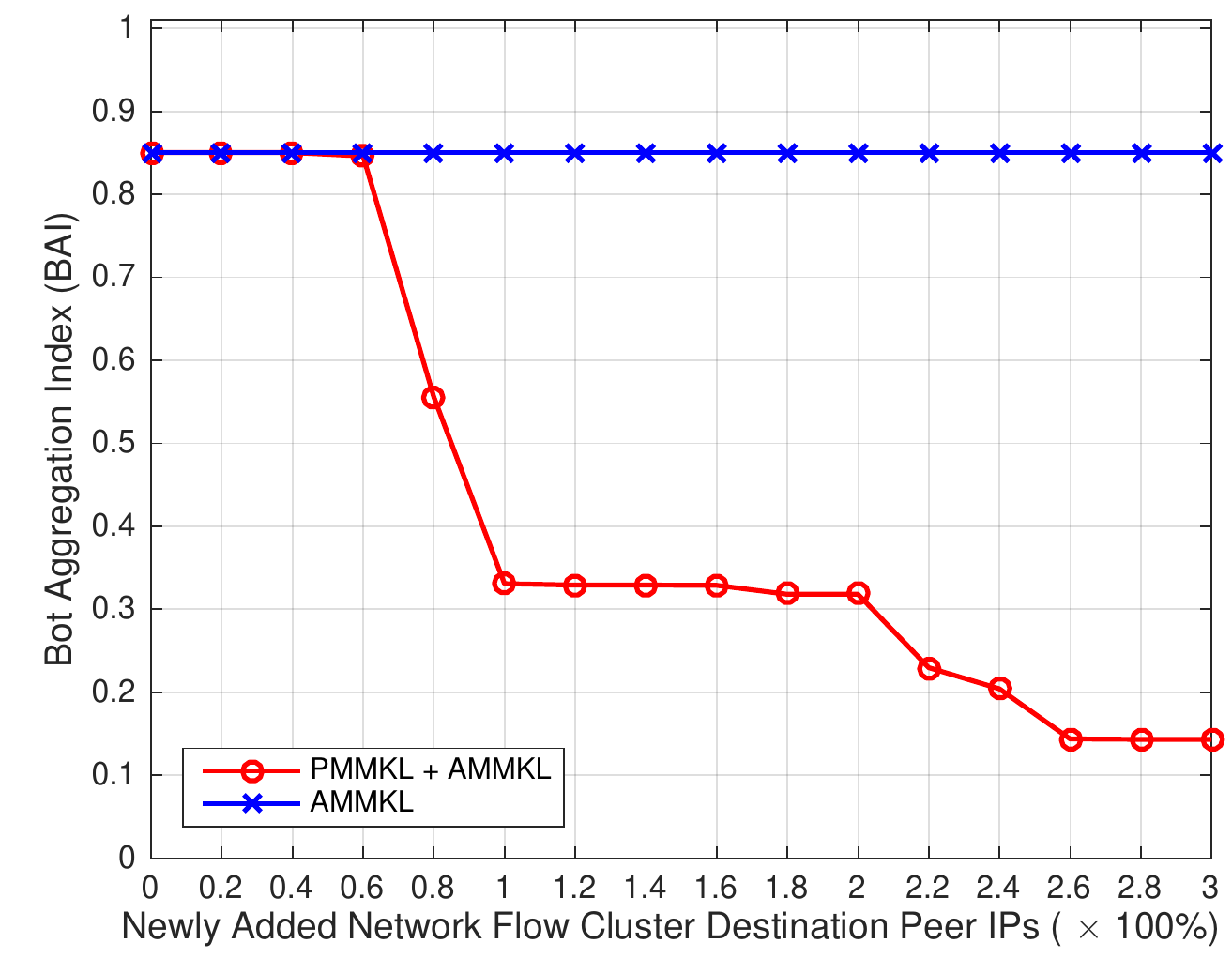}
                \caption{}
                \label{fig:ATT2_BAI}
        \end{subfigure}
        ~ %add desired spacing between images, e. g. ~, \quad, \qquad, \hfill etc.
        \begin{subfigure}[b]{0.32\textwidth}
                \includegraphics[width=\textwidth]{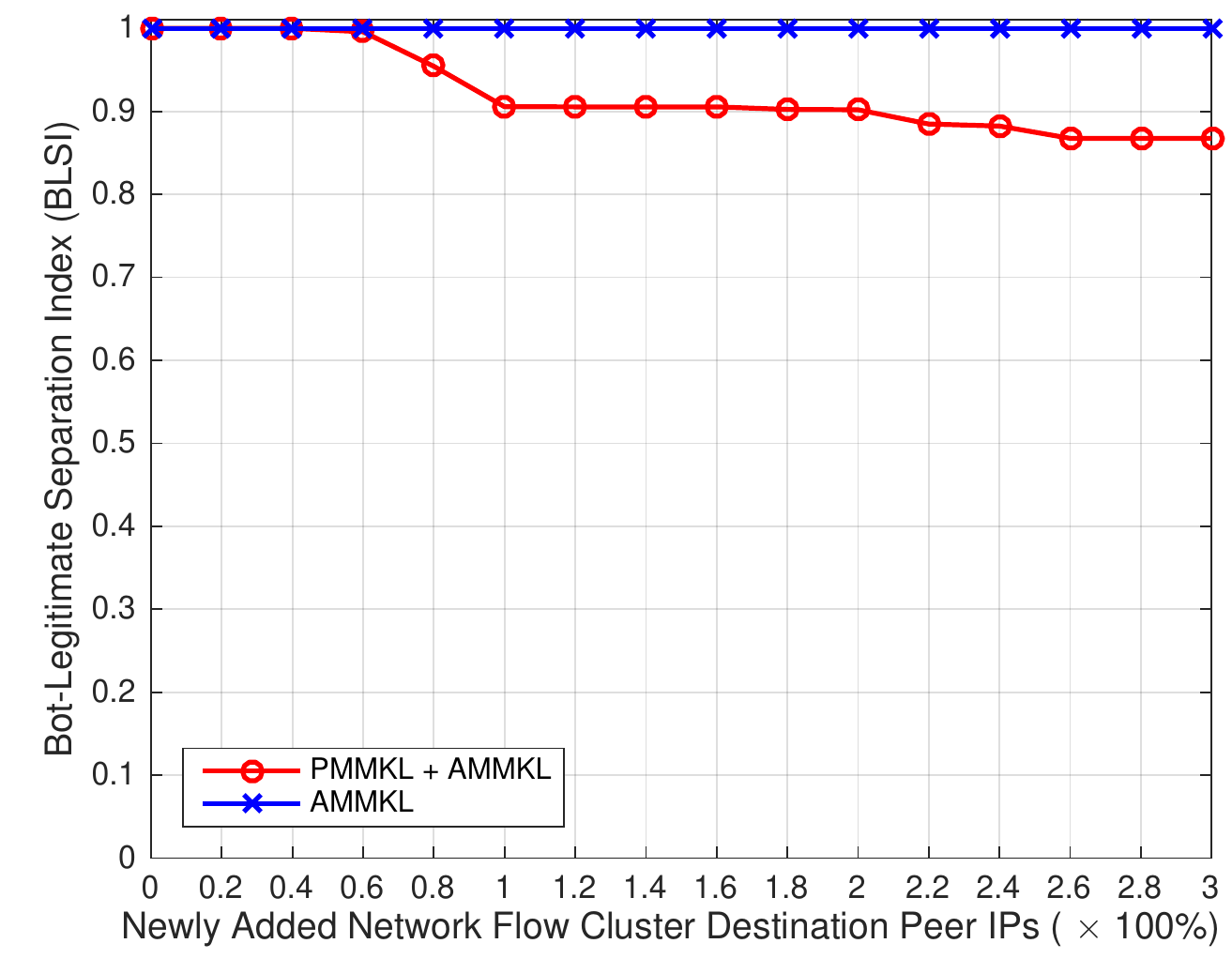}
                \caption{}
                \label{fig:ATT2_BLSI}
        \end{subfigure}
        \caption{The community detection results when conducting AMMKL, and when combining PMMKL and AMMKL. (a) Bot Separation Index (BSI). (b) Bot Aggregation Index (BAI). (c) Bot-Legitimate Separation Index (BLSI).}
        \label{fig:ATT2_CD}
\end{figure*}

To simulate this attack on each ED, after the P2P network flow detection procedure, for each botnet network flow cluster that communicates with $n$ peers, we inserted certain network flows communicating with an extra of $\gamma * n$ randomly selected peers.
As shown in Fig.~\ref{fig:ATT2_CD}, our community detection component is robust to AMMKL, since both BAI and BLSI were unchanged and only BSI dropped a little bit when $\gamma$ increased. When combining both attacks, both BSI and BAI dropped a lot, and BLSI dropped from 1.0 to around 0.88, as $\gamma$ increasing from 0.0 to 3.0. This is because when combining both attacks, as $\gamma$ increasing, the community detection component tends to cluster different types of bots into the same community and separate the same type of bots into different communities. The good news is, it can still well separate bots and legitimate P2P hosts into different communities. In summary, even though combining both attacks makes it harder for our method to separate different or aggregate the same type of bots, Enhanced PeerHunter is still robust in separating P2P bots from other hosts in the community detection process.

\begin{figure*}[tb]
\captionsetup{font=footnotesize}
        \centering
        \begin{subfigure}[b]{0.32\textwidth}
                \includegraphics[width=\textwidth]{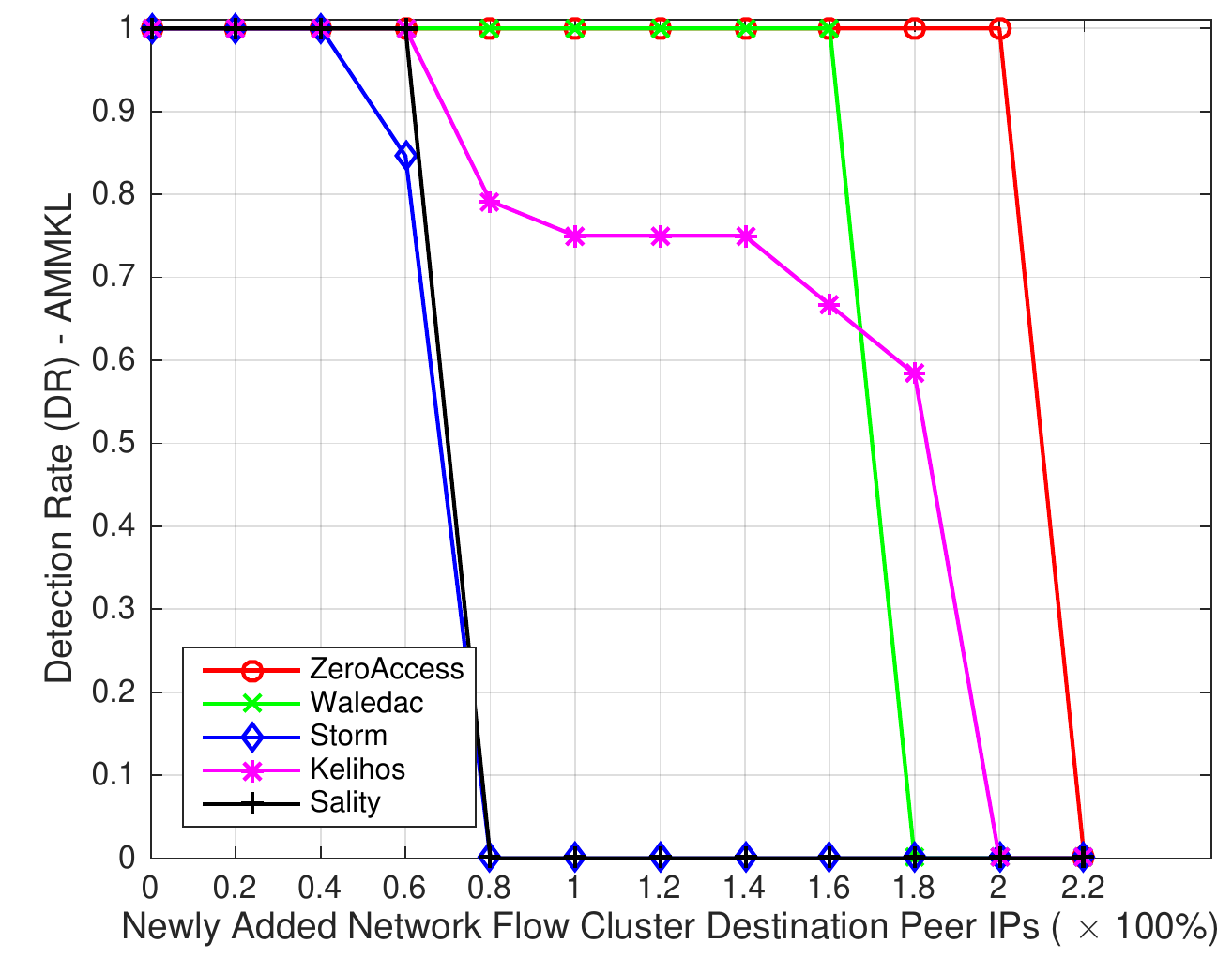}
                \caption{}
                \label{fig:ATT2_DR}
        \end{subfigure}%
        ~ %add desired spacing between images, e. g. ~, \quad, \qquad, \hfill etc.
        \begin{subfigure}[b]{0.32\textwidth}
                \includegraphics[width=\textwidth]{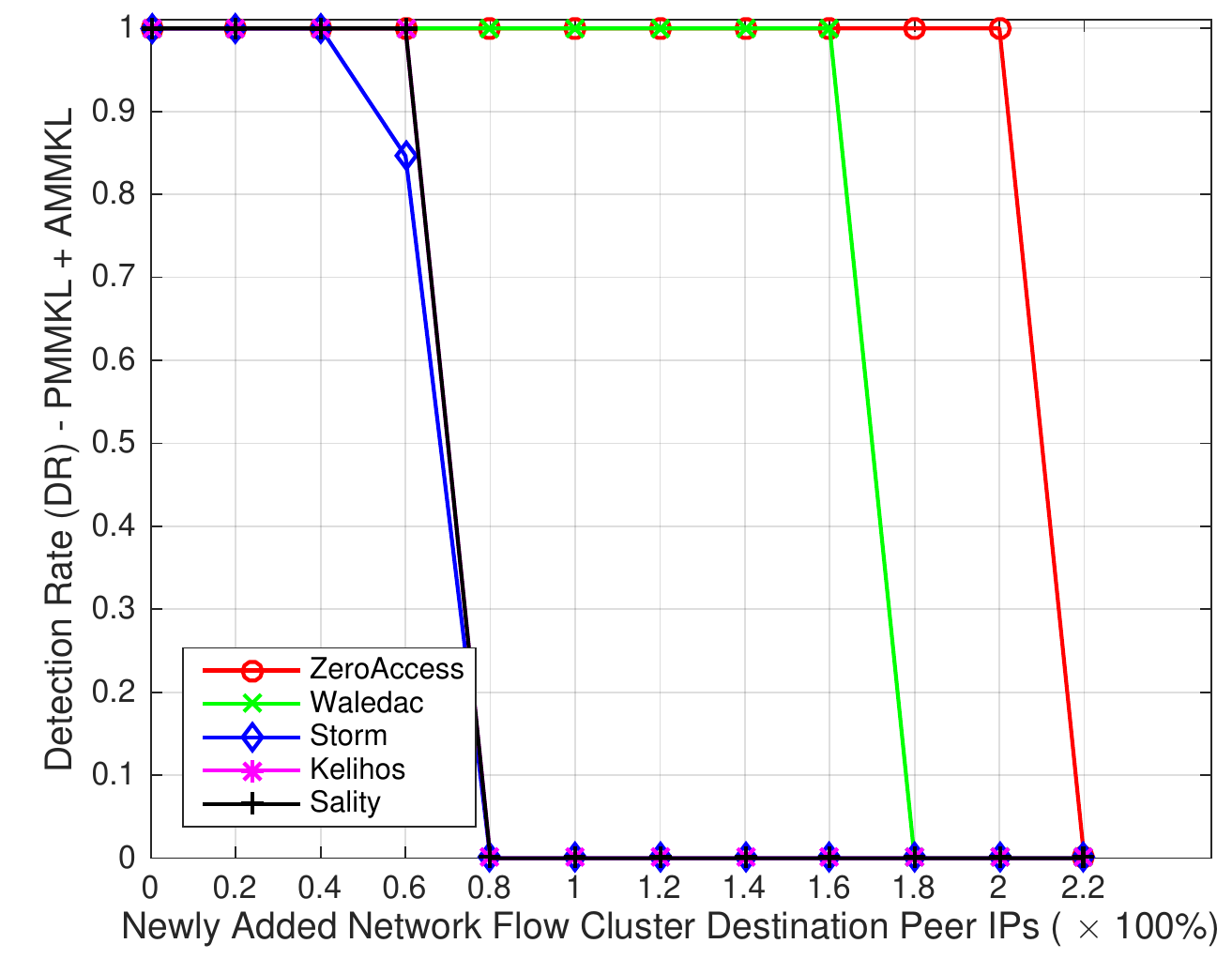}
                \caption{}
                \label{fig:ATT2_ATT1_DR}
        \end{subfigure}
        ~ %add desired spacing between images, e. g. ~, \quad, \qquad, \hfill etc.
        \begin{subfigure}[b]{0.32\textwidth}
                \includegraphics[width=\textwidth]{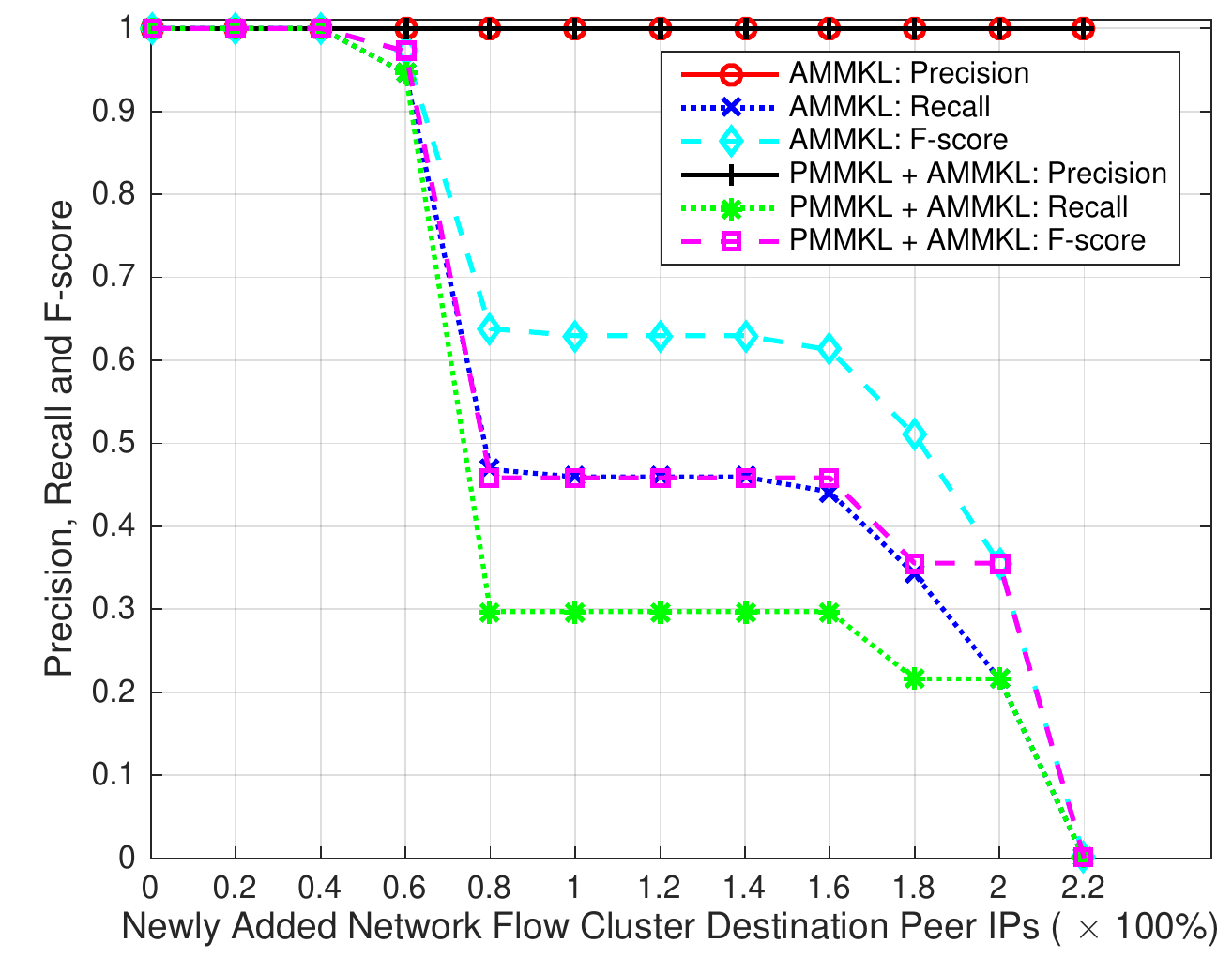}
                \caption{}
                \label{fig:ATT2_Other}
        \end{subfigure}
        \caption{The P2P botnet detection results. (a) P2P botnet detection rate when conducting AMMKL. (b) P2P botnet detection rate when combining PMMKL and AMMKL. (c) Precision, recall and F-score, when conducting AMMKL, and when combining PMMKL and AMMKL.}
        \label{fig:ATT2_BD}
\end{figure*}

As shown in Fig.~\ref{fig:ATT2_Other}, both scenarios (i.e., AMMKL and combining both attacks) did not introduce new false positives (i.e., precisions equals to 1.0). Compared with conducting AMMKL, combining both attacks has more influences on the dropping of detection rate. Fig.~\ref{fig:ATT2_DR} and Fig.~\ref{fig:ATT2_ATT1_DR} illustrate the detection rate of each botnet under two different scenarios, where the detection rate of different botnets started to drop around different $\gamma$. Table~\ref{table:ATT2_Evade} shows the analysis of all 5 botnets. Take Storm for instance, to affect the detection of Storm, each P2P network flow cluster of Storm needs to communicate with at least an extra 40\% of its current peers, and in order to completely evade our system, $\gamma$ needs to be increased to 80\%. Consider the fact that each Storm host generates an average of 67 P2P network flow clusters in 24 hours, and each network flow cluster communicates to an average of 740 peers. As such, to completely evade our system, each Storm host must communicate with at least an extra of $67 \times 740 \times 80\% \approx 39,664$ peers. In this case, it makes the P2P botnet less stealthy, less efficient and more exposed to trigger anomaly-based P2P botnet detection approaches \cite{feily2009survey}.
In conclusion, although our system could not completely mitigate AMMKL, conducting AMMKL makes the botnets less stealthy, less efficient and more exposed, which still shows a winning of our system against P2P botnets.

\begin{table*}[!t]
\footnotesize
\captionsetup{font=footnotesize}
\caption{Effort needed for different P2P botnets to completely evade Enhanced PeerHunter under AMMKL.}
\label{table:ATT2_Evade}
\centering
\begin{tabular}{c|c|c|c|c|c}
\hline
 - & \bfseries \# of P2P flow clusters & \bfseries \# of peers per flow cluster & \bfseries \# of peers per host & \bfseries $\gamma$ & \bfseries extra \# of peers needed \\
\hline
  ZeroAccess  & 3 & 686 & 2,058 & 220\% & 4,528\\
  \hline
  Waledac  & 171 & 244 & 41,724 & 180\% & 75,104\\
  \hline
  Storm  & 67 & 740 & 49,580 & 80\% & 39,664\\
  \hline
  Kelihos  & 15 & 252 & 3,780 & 200\% & 7,560\\
  \hline
  Sality  & 1,158 & 918 & 1,063,044 & 80\% & 850,436\\
\hline
\end{tabular}
\end{table*}

\subsection{Comparison to Zhang \textit{et al.} \cite{zhang2014building}}
We compared our system to one of the state of art P2P botnet detection system Zhang \textit{et al.} \cite{zhang2014building}. They proposed a scalable botnet detection system capable of detecting stealthy P2P botnets (i.e., in the waiting stage), where no knowledge of existing malicious behavior is required in advance. The system first applies a two-step flow clustering approach to create the fingerprints of hosts that have engaged in P2P activities. Afterwards, it applies two layers of filtering to detect potential P2P bots: a coarse-grained filtering to detect ``persistent'' P2P hosts that have longer active time of P2P behaviors, and a fine-grained filtering that applies hierarchical clustering to group pairs of P2P hosts that have less distance between their fingerprints. Our system shares many similarities with Zhang \textit{et al.} \cite{zhang2014building}. For instance, both systems are (a) using network flow-based approach, (b) using unsupervised approach (i.e., no knowledge of existing malicious behaviors are required and have the potential to detect unknown botnets), (c) claiming to work while the botnet traffic are overlapped with the legitimate P2P traffic on the same set of hosts, (d) designed to have the built-in scalability, and (e) deployed at the network boundary (e.g., gateway), thus could be evaluated on the same datasets.

The main differences between our system and Zhang \textit{et al.} \cite{zhang2014building} are listed as follows.
First, two systems are using different network flow features. Zhang \textit{et al.} \cite{zhang2014building} uses the absolute number of bytes and packets of each flow; Enhanced PeerHunter uses the bytes-per-packet rate of each flow.
Second, two systems are using different approach to cluster network flows (i.e., at different granularity). Zhang \textit{et al.} \cite{zhang2014building} uses a two-step distance-based clustering (i.e., k-means, BIRCH) to cluster network flows of similar feature values; Enhanced PeerHunter clusters the network flows that have exactly the same feature values.
Third, two systems apply the botnet detection step at different levels (i.e., host-level or network-flow-level). Zhang \textit{et al.} \cite{zhang2014building} uses the distance between each pair of hosts to detect bots; Enhanced PeerHunter uses the distance between each pair of network flows to detect botnet network flow communities and then further identify the corresponding bots.
Last but not least, two systems are using different heuristics to detect botnets. Zhang \textit{et al.} \cite{zhang2014building} uses an threshold on the height of the hierarchical clustering dendrogram to detect bot clusters, which is very sensitive to the experimental datasets (as shown in Table~\ref{table:Eva_ATT1_BD}); Enhanced PeerHunter uses network-flow level community behavior analysis (i.e., AVGDDR and AVGMCR) to identify botnet (network flow) communities, which is more robust to the proposed attacks and can also be extended to other/new community behaviors.

We implemented a prototype system of Zhang \textit{et al.} \cite{zhang2014building}, since Zhang \textit{et al.} \cite{zhang2014building} did not have a publicly available implementation. Most of our implementations followed the description as in \cite{zhang2014building}, other than the system parallelization, which has no impact on the system effectiveness evaluation. The experimental datasets used in both works are also different. For instance, we evaluated our system on 100 synthetic experimental datasets (of different background traffic and different topology, as described in Section~\ref{sec:sec5_1_3}) and took the average results; Zhang \textit{et al.} \cite{zhang2014building} was evaluated on single customized dataset. Furthermore, even though both datasets use the same 24 hours time window, our datasets have much more internal hosts (i.e., 10,000 vs. 953), higher legitimate P2P hosts to P2P bots ratio (i.e., 727:37 vs. 8:16), and more types of botnets (i.e., 5 vs. 2). To summarize, our experimental datasets is more challenging and comprehensive.

We applied our implemented Zhang \textit{et al.} \cite{zhang2014building} on the same experimental datasets as Enhanced PeerHunter under two circumstances (i.e., No Attack and PMMKL). We followed the same settings for most of the system parameters as described in \cite{zhang2014building}, such as $\Theta_{BGP}=50$, $\Theta_{p2p}=0.5$, $K=4,000$, $\lambda=0.5$. Since the default value of $\Theta_{bot}$ (i.e., 0.95) used by the original paper, did not perform well on our dataset, we evaluated Zhang \textit{et al.} \cite{zhang2014building} using two other different well selected values of $\Theta_{bot}$ (i.e., 0.6 and 0.8) that shows better results.

From the experimental results (Table~\ref{table:Eva_ATT1_BD}), we achieved several observations as follows.
First, Zhang \textit{et al.} \cite{zhang2014building} is more sensitive to the experimental dataset. For instance, Zhang \textit{et al.} \cite{zhang2014building} was reported to achieve 100\% detection rate and 0.2\% false positive rate on their own datasets (using $\Theta_{bot}=0.95$), while could not achieve similar results on our datasets using either the default parameter ($\Theta_{bot}=0.95$) or the well selected parameter ($\Theta_{bot}=0.6$ or $\Theta_{bot}=0.8$).
Second, as discussed in Section~\ref{sec:sec5_5}, our system is more stable and effective over a large range of system parameters ($\Theta_{avgddr}$ and $\Theta_{avgmcr}$), while Zhang \textit{et al.} \cite{zhang2014building} is more sensitive to its system parameter ($\Theta_{bot}$). For instance, Zhang \textit{et al.} \cite{zhang2014building} had higher precision (lower false positives) and lower recall (higher false negatives) while using $\Theta_{bot}=0.6$ comparing with using $\Theta_{bot}=0.8$.
Third, our system outperforms Zhang \textit{et al.} \cite{zhang2014building} in terms of the detection rate of different botnets, the overall precision, recall and false positives. For instance, our system achieved 100\% detection rate with zero false positives under different circumstances, while Zhang \textit{et al.} \cite{zhang2014building} failed to detect all the bots under both well selected parameters. At last, our system is more robust to PMMKL attack. For instance, PMMKL attack had no impact on the effectiveness of our system, while decreasing the F-score of Zhang \textit{et al.} \cite{zhang2014building} from 97.3\% to 75\% ($\Theta_{bot}=0.6$) or from 74.6\% to 72.7\% ($\Theta_{bot}=0.8$).

\section{Discussion}
\label{sec:sec8}
\subsection{Evasions and Possible Solutions}
To avoid being detected by Enhanced PeerHunter, the botmaster could use a combination of the following three approaches: (a) adding randomized paddings or junk packets to influence the bytes-per-packet characteristics for network flow clustering, (b) reducing the number or rate of destination diversity, or (c) reducing the number or rate of mutual contacts. To deal with the randomized spatial-communication behavior, we could adopt more time-communication features, such as packet/flow duration and inter-packet delays, or apply more generalized features, such as the distribution, mean or standard deviation of bytes-per-packet. The other two evasion approaches would be the victory of our system. On one hand, to reduce the number or rate of destination diversity, a bot has to limit its communication to the network of certain locations, which degrades the P2P botnet into a centralized fashion. On the other hand, reducing the number of mutual contacts means there will be less bots targeting on the same set of victims, and less bots playing the role as botmasters, which will jeopardize the effectiveness and the decentralized structure of a P2P botnet. Also, as shown in Section~\ref{sec:sec5_6_2}, reducing the rate of mutual contacts while maintaining the same number of mutual contacts (i.e., by conducting AMMKL) will make the botnets less stealthy, less efficient and more exposed to the other detection systems (e.g., anomaly-based botnet detection using high volumes of network traffic).

\subsection{The deployment of Enhanced PeerHunter}
In the previous sections, we simply assumed that our system is deployed at the boundary of a single organization. In this section, we discuss about the deployment of Enhanced PeerHunter in three more realistic scenarios.
\subsubsection{The number of bots within an organization is too small} It would be challenging to build the MCG of botnet communities (i.e., the number of bots belonging to the same botnet is less than 3). In this case, we can deploy multiple Enhanced PeerHunter systems at the boundaries of multiple organizations, and correlate the network flows collected by those multiple Enhanced PeerHunter systems to build an appropriate size of MCG to detect botnet communities.
\subsubsection{The number of bots within an organization is too large} The mutual contacts of certain bots might be within the organization internal network, hence invisible to the single system monitoring at the network boundary. In this case, we can deploy multiple Enhanced PeerHunter systems within the organization, that divide the organization network into several appropriate size of sub-internal networks. Each system is responsible for one sub-internal network.
\subsubsection{The botmaster knows the system deployment location} In this way, the botmaster could assign the location of bots or control the communications of the bots based on the knowledge of the system deployment location to evade our system. For instance, the botmaster could assign bots into different sub-internal networks, and instruct most of the bots communicate with the others within the same sub-internal network. In this case, we could use the concept and idea of Moving Target Defense (MTD) \cite{albanese2018defending} to develop a strategy that makes it more difficult for botmasters to learn the deployment locations of our systems, by dynamically changing the settings or deployments of our systems.

\subsection{Extend Enhanced PeerHunter to detect other botnets}
Although Enhanced PeerHunter is designed to detect P2P botnets, %as shown in Section~\ref{sec:sec5_5_4},
our idea of using mutual contacts graph has the potential to detect not only unknown botnets, but also the other types of botnets (e.g., centralized botnets, such as IRC botnets \cite{ma2010novel}, mobile botnets \cite{zhao2012cloud}). Since bots are usually controlled by machines, rather than humans, bots from the same botnets tend to communicate with a similar set of peers or attacking targets. For instance, bots from the same IRC botnets tend to contact a similar set of C\&C servers, while bots from the same mobile botnets tend to contact a similar set of satellite servers. Hence, we argue that Enhanced PeerHunter could be easily extended to detect the other types of botnets.

\section{Conclusion}
\label{sec:sec7}
We present a novel community behavior analysis based P2P botnet detection system, Enhanced PeerHunter, which operates under several challenges: (a) botnets are in their waiting stage; (b) the C\&C channel has been encrypted; (c) the botnet traffic are overlapped with legitimate P2P traffic on the same host; (d) no bot-blacklist or ``seeds'' are available; (e) none statistical traffic patterns known in advance; and (f) does not require to monitor individual host. We propose three types of community behaviors (i.e., flow statistical features, numerical community features and structural community features) that can be used to detect P2P botnets effectively. In the experimental evaluation, we propose a network traces sampling and mixing method to make the experiments as unbiased and challenging as possible. Experiments and analysis were conducted to show the effectiveness and scalability of our system. With the best parameter settings, our system achieved 100\% detection rate with none false positives. We also propose two mimicking legitimate P2P application attacks (i.e., PMMKL and AMMKL). The experiment results showed that our system is robust to PMMKL, and will make the botnets less stealthy, less efficient and more exposed while conducting AMMKL.

\ifCLASSOPTIONcaptionsoff
  \newpage
\fi

\bibliographystyle{IEEEtran}
\bibliography{IEEEabrv,tifs_18_dz}
\vskip 0pt plus -1fil
\begin{IEEEbiography}[{\includegraphics[width=1.05in,height=2.0in,clip,keepaspectratio]{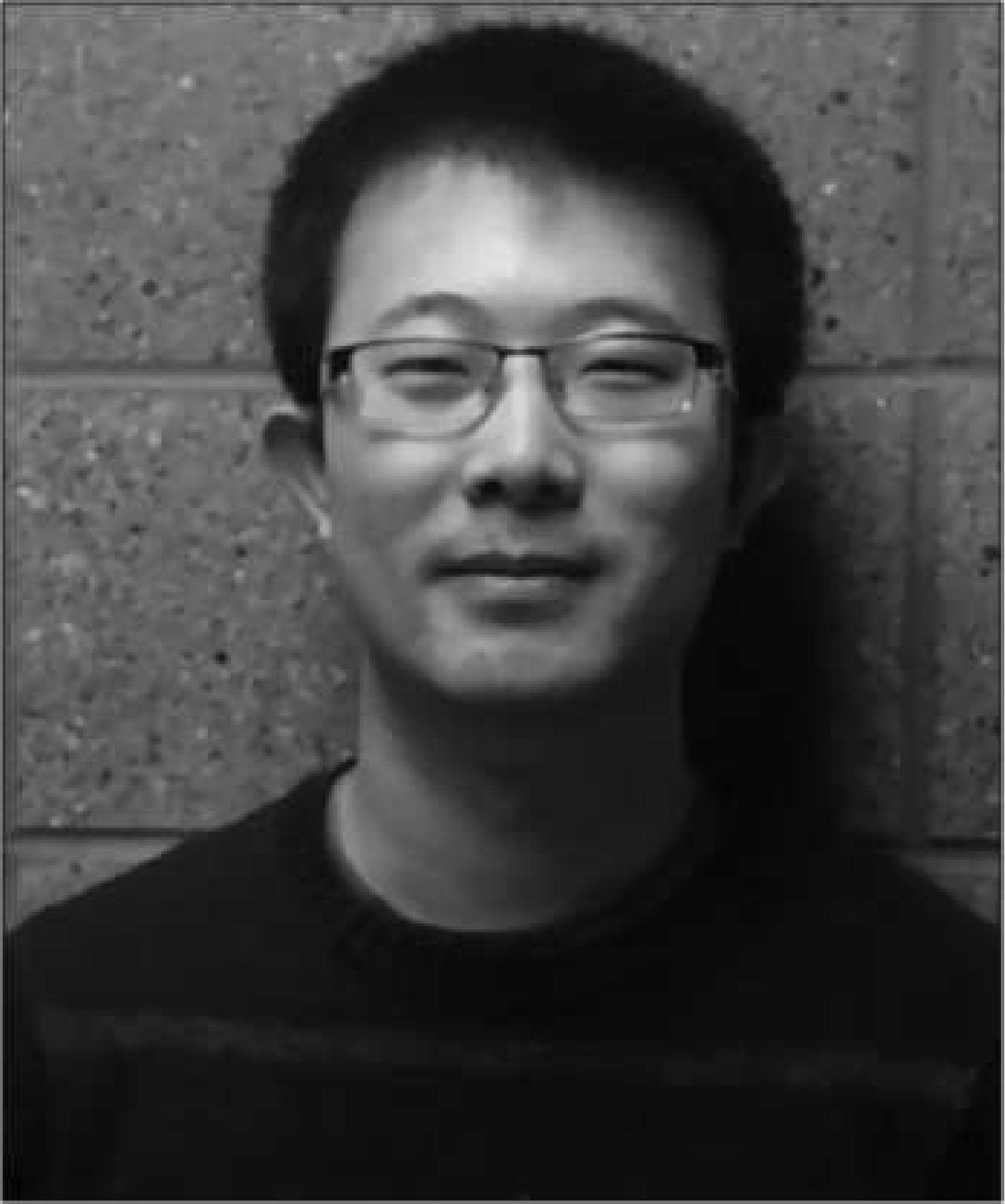}}]{Di Zhuang}
(S'15) received the B.E. degree in computer science and information security from Nankai University, China. He is currently pursuing the Ph.D. degree in electrical engineering with University of South Florida, Tampa. His research interests include cyber security, social network science, privacy enhancing technologies, machine learning and big data analytics. He is a student member of IEEE.
\end{IEEEbiography}
\vskip 0pt plus -1fil
\begin{IEEEbiography}[{\includegraphics[width=1.05in,height=2.0in,clip,keepaspectratio]{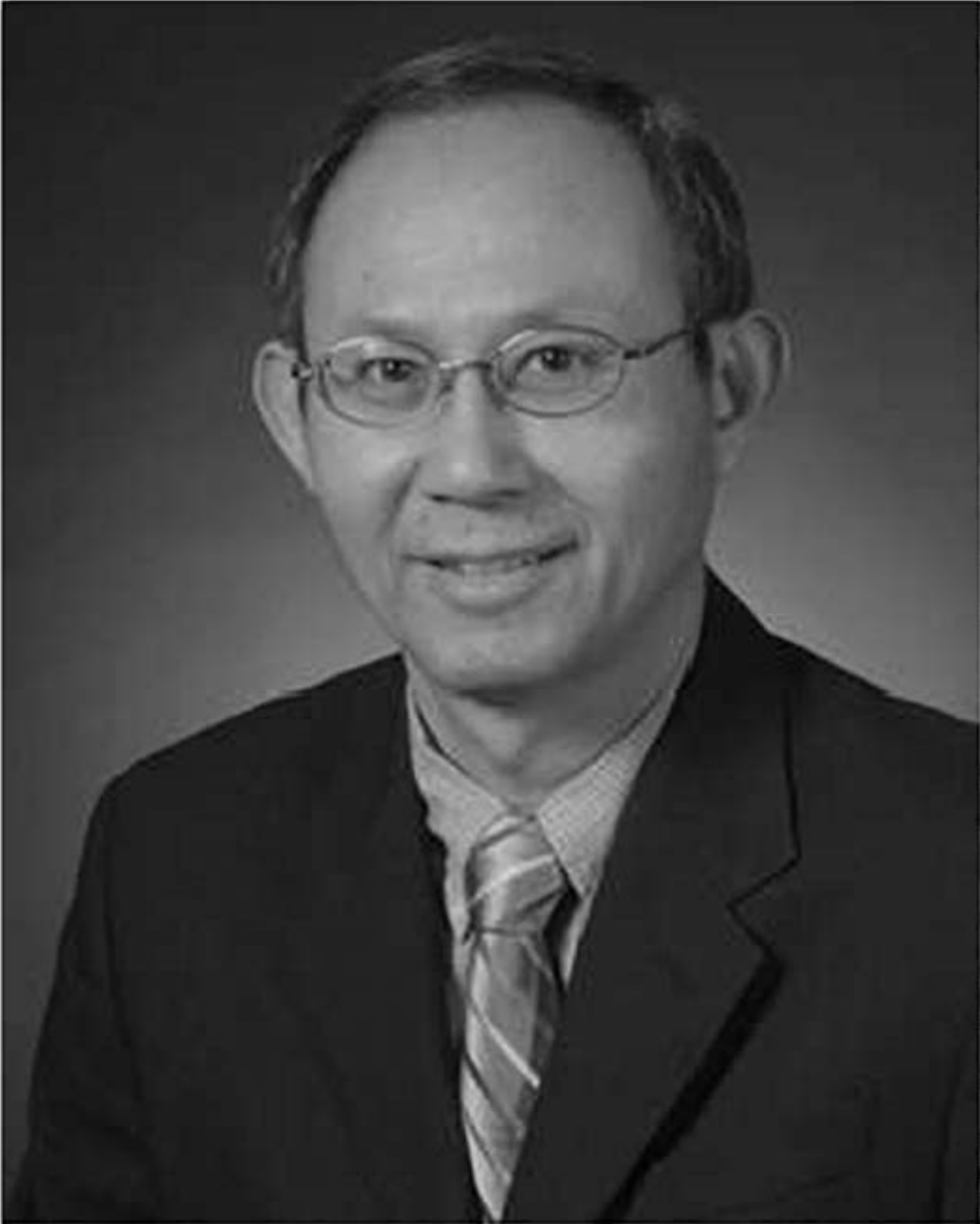}}]{J. Morris Chang}
(SM'08) is a professor in the Department of Electrical Engineering at the University of South Florida. He received the Ph.D. degree from the North Carolina State University. His past industrial experiences include positions at Texas Instruments, Microelectronic Center of North Carolina and AT\&T Bell Labs. He received the University Excellence in Teaching Award at Illinois Institute of Technology in 1999. His research interests include: cyber security, wireless networks, and energy efficient computer systems. In the last six years, his research projects on cyber security have been funded by DARPA. Currently, he is leading a DARPA project under Brandeis program focusing on privacy-preserving computation over Internet. He is a handling editor of Journal of Microprocessors and Microsystems and an editor of IEEE IT Professional. He is a senior member of IEEE.
\end{IEEEbiography}
\end{document}